\documentclass[fleqn,usenatbib]{mnras}

\usepackage{newtxtext,newtxmath}

\usepackage[T1]{fontenc}
\usepackage{ae,aecompl}

\usepackage{graphicx}	
\usepackage{amsmath}	
\usepackage{amssymb}	

\usepackage{xspace}
\usepackage{soul}

\newcommand{\epic}{K2-137\xspace}
\newcommand{\ktwo}{{\it K2}\xspace}
\newcommand{\kep}{\textit{Kepler}\xspace}
\newcommand{\feh}{\mbox{[Fe/H]}\xspace}
\newcommand{\teff}{\mbox{$T_{\rm *, eff}$}\xspace}
\newcommand{\logg}{\mbox{$\log g_*$}\xspace}
\newcommand{\kms}{\mbox{km\,s$^{-1}$}\xspace}
\newcommand{\ms}{\mbox{m\,s$^{-1}$}\xspace}
\newcommand{\mplanet}{\mbox{$M_{\rm p}$}\xspace}
\newcommand{\rplanet}{\mbox{$R_{\rm p}$}\xspace}
\newcommand{\mjup}{\mbox{$\mathrm{M_{\rm Jup}}$}\xspace}

\newcommand{\me}{\mbox{$\mathrm{M_{\rm \oplus}}$}\xspace}
\newcommand{\re}{\mbox{$\mathrm{R_{\rm \oplus}}$}\xspace}
\newcommand{\mstar}{\mbox{$M_{*}$}\xspace}
\newcommand{\rstar}{\mbox{$R_{*}$}\xspace}
\newcommand{\densstar}{\mbox{$\rho_*$}\xspace}
\newcommand{\msol}{\mbox{$\mathrm{M_\odot}$}\xspace}
\newcommand{\rsol}{\mbox{$\mathrm{R_\odot}$}\xspace}

\newcommand{\lstar}{\mbox{$L_{*}$}\xspace}
\newcommand{\lsol}{\mbox{$\mathrm{L_\odot}$}\xspace}

\title[\epic~b]{\epic~b: an Earth-sized planet in a 4.3-hour orbit around an M-dwarf}

\author[A. M. S. Smith et al.]{
A. M. S. Smith,$^{1}$\thanks{E-mail: Alexis.Smith@dlr.de}
J.~Cabrera$^{1}$, 
Sz.~Csizmadia$^{1}$,
F.~Dai$^{2}$,
D.~Gandolfi$^{3}$, 
\newauthor
T.~Hirano$^{4}$, 
J.~N.~Winn$^{5}$,
S.~Albrecht$^{6}$,
R.~Alonso$^{7,8}$,
G.~Antoniciello$^{3}$,
\newauthor
O.~Barrag\'{a}n$^{3}$,
H.~Deeg$^{7,8}$, 
Ph.~Eigm\"uller$^{1}$, 
M.~Endl$^{9}$, 
A.~Erikson$^{1}$,
\newauthor
M.~Fridlund$^{10,11,7}$,
A.~Fukui$^{12}$,
S.~Grziwa$^{13}$, 
E.~W.~Guenther$^{14}$, 
A.~P.~Hatzes$^{14}$,
\newauthor
D.~Hidalgo$^{7,8}$,
A.~W.~Howard$^{15}$,
H.~Isaacson$^{16}$,
J.~Korth$^{13}$,
M.~Kuzuhara$^{17,18}$,
\newauthor
J.~Livingston$^{19}$,
N.~Narita$^{19,17,18}$,
D.~Nespral $^{7,8}$, 
G.~Nowak$^{7,8}$, 
E.~Palle$^{7,8}$,
\newauthor
M.~P\"atzold$^{13}$,
C.M.~Persson$^{11}$,   
E.~Petigura$^{20}$,
J.~Prieto-Arranz$^{7,8}$,
H.~Rauer$^{1,21}$,
\newauthor
I.~Ribas$^{22}$,
and V.~Van~Eylen$^{10}$
\\
$^{1}$Institute of Planetary Research, German Aerospace Center, Rutherfordstrasse 2, 12489 Berlin, Germany\\
$^{2}$Department of Physics and Kavli Institute for Astrophysics and Space Research, Massachusetts Institute of Technology, Cambridge, MA 02139,USA\\
$^{3}$Dipartimento di Fisica, Universit\'a di Torino, Via P. Giuria 1, I-10125, Torino, Italy\\
$^{4}$Department of Earth and Planetary Sciences, Tokyo Institute of Technology, 2-12-1 Ookayama, Meguro-ku, Tokyo 152-8551, Japan\\
$^{5}$Princeton University, Department of Astrophysical Sciences, 4 Ivy Lane, Princeton, NJ 08544 USA\\
$^{6}$Stellar Astrophysics Centre, Department of Physics and Astronomy, Aarhus University, Ny Munkegade 120, DK-8000 Aarhus C, Denmark\\
$^{7}$Instituto de Astrof\'\i sica de Canarias (IAC), 38205 La Laguna, Tenerife, Spain\\
$^{8}$Departamento de Astrof\'\i sica, Universidad de La Laguna (ULL), 38206 La Laguna, Tenerife, Spain\\
$^{9}$Department of Astronomy and McDonald Observatory, University of Texas at Austin, 2515 Speedway, Stop C1400, Austin, TX 78712, USA\\
$^{10}$Leiden Observatory, Leiden University, 2333CA Leiden, The Netherlands\\
$^{11}$Department of Earth and Space Sciences, Chalmers University of Technology, Onsala Space Observatory, 439 92, Onsala, Sweden\\
$^{12}$Okayama Astrophysical Observatory, National Astronomical Observatory of Japan, Asakuchi, 719-0232 Okayama, Japan\\
$^{13}$Rheinisches Institut f\"ur Umweltforschung an der Universit\"at zu K\"oln, Aachener Strasse 209, 50931 K\"oln, Germany\\
$^{14}$Th\"uringer Landessternwarte Tautenburg, Sternwarte 5, 07778 Tautenburg, Germany\\
$^{15}$Department of Astronomy, California Institute of Technology, Pasadena, CA, USA\\
$^{16}$Astronomy Department, University of California, Berkeley, CA, USA\\
$^{17}$Astrobiology Center, NINS, 2-21-1 Osawa, Mitaka, Tokyo 181-8588, Japan\\
$^{18}$National Astronomical Observatory of Japan, NINS, 2-21-1 Osawa, Mitaka, Tokyo 181-8588, Japan\\
$^{19}$Department of Astronomy, The University of Tokyo, 7-3-1 Hongo, Bunkyo-ku, Tokyo 113-0033, Japan\\
$^{20}$Geological and Planetary Sciences, California Institute of Technology, Pasadena, CA, US\\
$^{21}$Center for Astronomy and Astrophysics, TU Berlin, Hardenbergstr. 36, 10623 Berlin, Germany\\
$^{22}$Institut de Ci\`{e}ncies de l'Espai (CSIC-IEEC), Carrer de Can Magrans, Campus UAB, E-08193 Bellaterra, Spain
}

\date{Accepted 2017 November 03. Received 2017 November 03; in original form 2017 July 14}

\pubyear{2017}

\begin{document}
\label{firstpage}
\pagerange{\pageref{firstpage}--\pageref{lastpage}}
\maketitle

\begin{abstract}

We report the discovery in \ktwo's Campaign 10 of a transiting terrestrial planet in an ultra-short-period orbit around an M3-dwarf. \epic~b completes an orbit in only 4.3~hours, the second-shortest orbital period of any known planet, just 4~minutes longer than that of KOI~1843.03, which also orbits an M-dwarf. Using a combination of archival images, AO imaging, RV measurements, and light curve modelling, we show that no plausible eclipsing binary scenario can explain the \ktwo light curve, and thus confirm the planetary nature of the system. The planet, whose radius we determine to be $0.89 \pm 0.09$~\re, and which must have a iron mass fraction greater than 0.45, orbits a star of mass $0.463 \pm 0.052$~\msol and radius $0.442 \pm 0.044$~\rsol.

\end{abstract}




\section{Introduction}

Numerous planets with very short orbital periods have been discovered over the last few years, including a number which complete an orbit of their host stars in less than 1 day. These include hot Jupiters (HJs) such as WASP-43b \citep{w43} and rocky planets such as CoRoT-7b \citep{corot7_1,corot7_2} and Kepler-78b \citep{kepler78,kepler78_2,kepler78_3}. This latter class has become known as the ultra-short-period (USP) planets. \cite{Sanchis-Ojeda14} found 106 USP planet candidates in the \kep data, and noted that the radius of such objects rarely exceeds 2~\re. They found that USP planets very often occur in multi-planet systems, with other planets in orbits shorter than 50~d. Several USP candidates have been detected by \ktwo, with \cite{Adams16} publishing a catalogue of 19 USP candidates from \ktwo campaigns 0--5.

The confirmed planet with the shortest orbital period is KOI~1843.03, which orbits with a period of 4.245~h. KOI~1843.03 was detected as a planetary candidate by \cite{Ofir_Dreizler_13}, and analysed more fully by \cite{KOI1843}. In this paper we report the discovery of \epic~b (= EPIC~228813918~b), a transiting planet, striking in its similarity to KOI~1843.03. \epic~b orbits a $V = 15.5$ M-dwarf in Virgo (Table~\ref{tab:stellar}) with a 4.3-h period. We use its \ktwo light curve, radial velocities, archival images, adaptive optics imaging and a catalogue of eclipsing binaries (EBs) to demonstrate that \epic is a transiting planetary system, and not an EB.

\begin{table}
\caption{Catalogue information for \epic}
\begin{center}
\begin{tabular}{ll} \hline
Parameter  & Value \\ \hline
RA (J2000.0) & 12h27m28.974s  \\
Dec (J2000.0) & $-06^{\circ}~11\arcmin~42.81\arcsec $\\
pmRA$^*$ (mas yr$^{-1}$) & $-82.6 \pm 3.7$ \\
pmDec$^*$ (mas yr$^{-1}$) & $-1.0 \pm 3.6$ \\
\hline
Magnitudes&(from EPIC$^\dagger$)\\
$B$&17.007 $\pm$ 0.090\\
$g\prime$&16.339 $\pm$ 0.040\\
$V$&15.498 $\pm$ 0.054\\
$r\prime$&14.955 $\pm$ 0.020\\
\kep&14.534\\
$i\prime$&13.760 $\pm$ 0.010\\
$J$ (2MASS)&11.764 $\pm$ 0.026\\
$H$ (2MASS)&11.126 $\pm$ 0.022\\
$K$ (2MASS)&10.882 $\pm$ 0.023\\
\hline
\multicolumn{2}{l}{Additional identifiers for \epic:}\\
\multicolumn{2}{l}{EPIC~228813918}\\
\multicolumn{2}{l}{UCAC 420-056244}\\
\multicolumn{2}{l}{2MASS J12272899-0611428}\\
\hline
\multicolumn{2}{l}{$^*$ Proper motions are from UCAC5}\\
\multicolumn{2}{l}{~~~~~\citep{UCAC5}}\\
\multicolumn{2}{l}{$\dagger$\ktwo's Ecliptic Plane Input Catalog}\\
\end{tabular}
\end{center}
\label{tab:stellar}
\end{table}

\section{Observations}
\label{sec:obs}
\subsection{\ktwo photometry}

\epic was observed as part of \ktwo's Campaign 10 (C10), from 2016 July 06 to 2016 September 20. The C10 field was repointed on 2016 July 13, and there was a 14~d gap in the observations from 2016 July 20 to 2016 August 03 due to a failure of module 4.

Using the {\sc DST} code of \cite{DST}, we searched the C10 light curves extracted by \cite{vburg} for periodic signals indicative of transiting exoplanets. A signal with a period of just 4.3~h and a depth of only $\sim 350$~ppm was detected in the light curve of \epic. The light curve was also searched for evidence of stellar rotation, or other activity, but none was found.

\subsection{Subaru/IRCS AO imaging and spectroscopy}
\label{sec:obs:ao}
We performed adaptive optics (AO) observations with the Infrared Camera and Spectrograph \citep[IRCS;][]{Subaru_IRCS} on the 8.2-m Subaru telescope at Mauna Kea, Hawaii, USA on UT 2017 May 22. For \epic, we conducted AO imaging to search for faint sources of contamination around the target, and also echelle spectroscopy with AO to obtain quasi-high resolution spectra ($R\sim 20\ 000$) for low-precision radial velocity (RV) measurements (Table~\ref{tab:rv}).

For the AO imaging, we adopted the high resolution mode (1 pix = 20 mas) with the $H-$band filter, and obtained two kinds of frames, one with saturation to search for faint sources and one without saturation to calibrate the flux. We repeated the 5-point dithering, resulting in total scientific exposures of 37.5~s and 750~s for saturated and unsaturated frames, respectively. 

Using echelle spectroscopy, we obtained $H-$band spectra (the standard $H^-$ setup) covering $1462-1838$ nm with ABBA-pattern nodding. During the night, we took three sets of echelle spectra covering a wide range of orbital phase, and the exposure times for each nodding position varied from 60~s to 200~s. We also obtained spectra of an A0 standard star (HIP 61318) for telluric correction, shortly after taking each set of target spectra.

\subsection{WIYN/NESSI speckle imaging}

We observed \epic with the NASA Exoplanet Star and Speckle Imager (NESSI) on the 3.5-m WIYN telescope at the Kitt Peak National Observatory, Arizona, USA on 2017 May 12. NESSI is a new instrument that uses high-speed electron-multiplying CCDs (EMCCDs) to capture sequences of 40~ms exposures simultaneously in two bands (Scott et al., in prep.). In addition to the target, we also observed nearby point source calibrator stars close in time to the science target. We conducted all observations in two bands simultaneously: a 'blue' band centered at 562nm with a width of 44nm, and a 'red' band centered at 832nm with a width of 40nm. The pixel scales of the 'blue' and 'red' EMCCDs are 0.01756 arcsec/pixel and 0.01819 arcsec/pixel, respectively.

\subsection{Keck/HIRES spectroscopy}
\label{sec:keck}
Two spectra were obtained with the HIRES instrument of the 10-m Keck I telescope on Mauna Kea, Hawaii, USA on 2017 May 13, covering a wavelength range of 364 -- 799~nm. The observations each used an exposure time of 600~s (giving a signal-to-noise ratio of 20 -- 25 per pixel), and were separated by 1.3~h such that the observations were within 20 -- 30 minutes of the predicted times of quadrature of the 4.3-hour orbit.

The observed spectra were cross-correlated with a number of template spectra; the best match was found to be a spectrum with $\teff = 3600$~K, and no evidence of a second star was found in the spectra. Radial velocities were determined using the telluric lines, and are reported in Table~\ref{tab:rv}.

\begin{table}
\caption{Radial velocity measurements of \epic}
\begin{center}
\begin{tabular}{llll}\hline
$\mathrm{BJD_{TDB}}$ & RV & $\sigma_{\mathrm{RV}}$ & Telescope/ \\
 $-2450000$ & \kms & \kms & Instrument \\
\hline
7886.898279  &  $-13.3$      &  0.1 & Keck/HIRES \\
7886.954710  &  $-13.4$      &  0.1 & Keck/HIRES \\
7895.448830  &  $-13.1$    &  0.1 & NOT/FIES \\
7896.428038  &  $-13.15$    &  0.05 & NOT/FIES \\
7896.795890  &  $-13.4$      &  1.4 & Subaru/IRCS \\
7896.801710  &  $-13.7$      &  0.8 & Subaru/IRCS \\
7896.870920  &  $-12.0$      &  0.6 & Subaru/IRCS \\
\hline
\\
\end{tabular}
\end{center}
\label{tab:rv}
\end{table}

\subsection{NOT/FIES spectroscopy}

We also acquired two high-resolution spectra of \epic with the FIbre-fed \'Echelle Spectrograph \citep[FIES;][]{Frandsen1999,Telting2014} mounted at the 2.56-m Nordic Optical Telescope (NOT) of Roque de los Muchachos Observatory (La Palma, Spain). We used the \emph{med-res} fibre, which provides a resolving power of $R=47\,000$ in the wavelength range 364\,--\,885~nm. The observations were carried out on UT 2017 May 21 and 22 as part of the observing program P55-019. We set the exposure time to 0.5 and 1.5 hours, leading to a signal-to-noise ratio per pixel at 650~nm of $\sim$15 and 25, respectively. We traced the intra-exposure RV drift of the instrument by acquiring long-exposed ($\sim$30 sec) ThAr spectra immediately before and after the target observations \citep{Gandolfi2015}. The data were reduced using standard IRAF and IDL routines. The RV measurements were extracted via multi-order cross-correlations with a template spectrum of the M2\,V star GJ\,411, that we observed in June 2017 with the same instrument set-up as \epic. The FIES RVs are listed in Table~\ref{tab:rv} along with their 1$\sigma$ uncertainties.

\subsection{NOT/ALFOSC spectroscopy}

We acquired a low-resolution spectrum of \epic with the Andalucia Faint Object Spectrograph and Camera (ALFOSC) mounted at the NOT. The observations were performed on UT 2017 May 22 as part of the same observing program as the FIES observations, under good and stable weather conditions, with seeing typically ranging between 1.0\arcsec and 1.3\arcsec. We used the Grism~\#\,4 coupled with a 0.5\arcsec-wide slit, leading to a resolving power of $R\,\approx\,700$ in the wavelength range 350 -- 950~nm. We removed cosmic-ray hits by combining three consecutive spectra of 720~s each. The data were reduced using standard IRAF routines.  Relative flux calibration was achieved observing the spectrophotometric standard star BD+26\,2606 from the compilation of \citet{Oke90}.  The signal-to-noise ratio of the extracted spectrum is $\sim$130 per pixel at 650~nm.

\section{Stellar characterisation}
\label{sec:star}

\begin{figure}
\begin{center}
\includegraphics[width=\columnwidth]{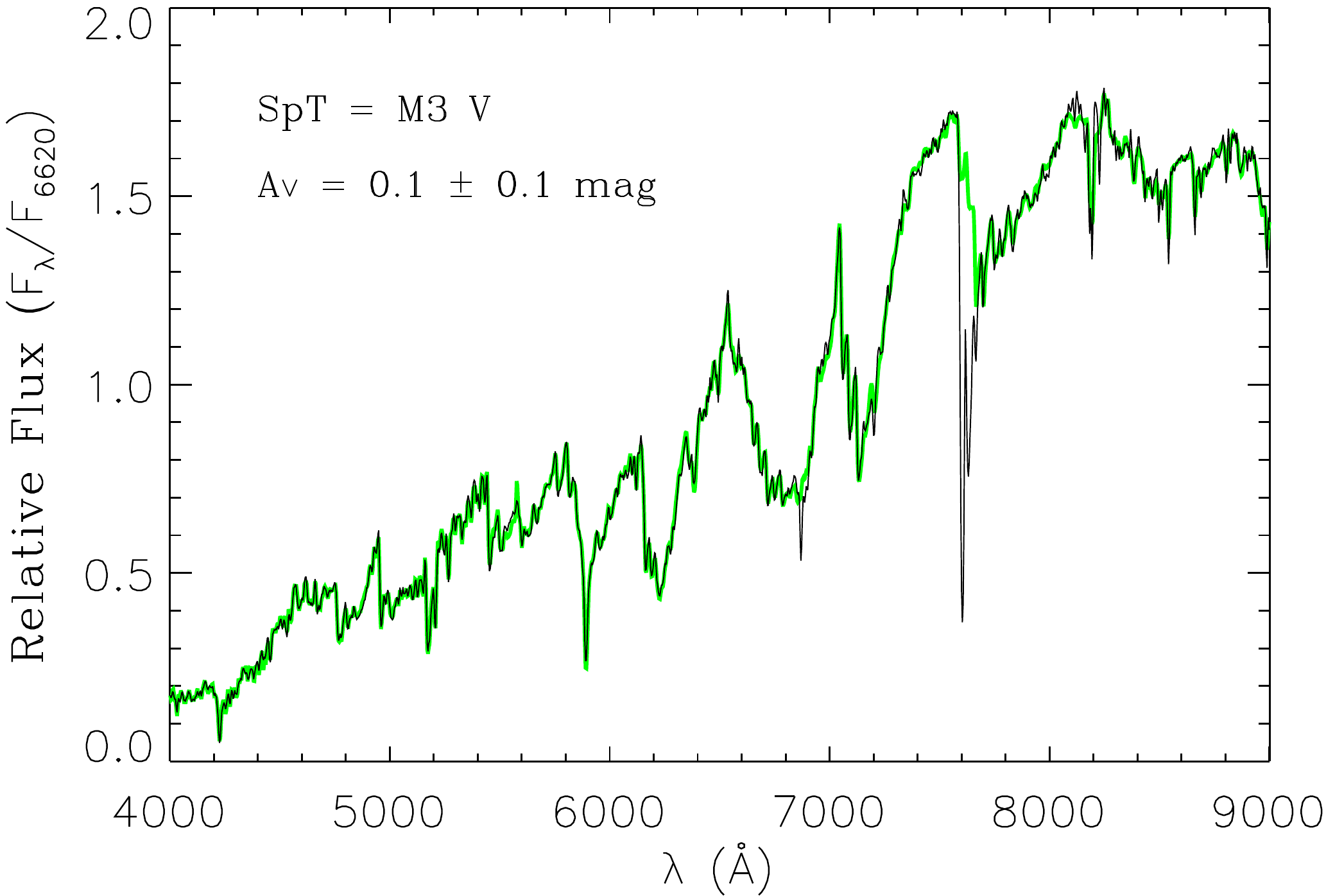}
\caption{Low-resolution ($R\,\approx\,700$) ALFOSC spectrum of \epic (thin black line). The best-fitting M3\,V template is overplotted with a thick green line. Both spectra are arbitrarily normalized to the flux at 662~nm.}. 
\label{fig:alfosc_spec}
\end{center}
\end{figure}

Following the method described in \citet{Gandolfi2008}, we derived the spectral type, luminosity class, and interstellar extinction ($A_\mathrm{V}$) of \epic from the low-resolution ALFOSC spectrum. Briefly, we fitted the observed data with a grid of M-type template spectra extracted from the public libraries of \citet{Martin1999, Hawley2002, LeBorgne2003, Bochanski2007}. Since both the ALFOSC and template spectra are flux calibrated, we accounted for the amount of reddening along the line-of-sight to the star assuming a normal value for the total-to-selective extinction ($R_\mathrm{V}$\,=\,$A_\mathrm{V}$/$E(B-V)$\,=\,3.1). We found that \epic is an M3\,V star with a low reddening consistent with zero $A_\mathrm{V}$\,=\,$0.1\pm0.1$~mag, as expected for a relatively nearby star within 100~pc of the Sun. The ALFOSC spectrum of \epic and the best fitting M3\,V stellar template are shown in Fig.~\ref{fig:alfosc_spec}.

The two Keck / HIRES spectra were analysed separately using {\sc SpecMatch-emp} \citep{Specmatch-emp}, to determine the stellar effective temperature, \teff, the stellar radius, \rstar\footnote{\rstar is reported by {\sc SpecMatch-emp} instead of the more commonly-used \logg, since \rstar has been directly measured for late-type library stars, and transformation to \logg would necessitate the use of stellar models with known systematic errors.}, and the stellar metallicity, \feh. {\sc SpecMatch-emp} parametrizes stellar spectra, by comparing them to a library of well-characterised stars. This stellar library contains 404 stars which have high-resolution ($R \approx 60\,000$) Keck/HIRES spectra, as well as properties derived from other observations (interferometry, asteroseismology, spectrophotometry) and from LTE spectral synthesis. The library covers stars of spectral types F1 -- M5 ($\teff \approx 3000 - 7000$~K, $\rstar \approx 0.1 - 16$~\rstar); \epic falls well within the covered range.

In validating {\sc SpecMatch-emp}, \cite{Specmatch-emp} found its performance to be better for late-type stars (\teff < 4500~K), where typical uncertainties are 70~K in \teff, ten percent in stellar radius, and 0.12 dex in \feh, justifying its use in our context. The resulting stellar parameters derived from each spectrum are in very good agreement with each other ($\ll$~1-$\sigma$). We adopt the mean of the two values, and list them in Table~\ref{tab:spec}.

Monte Carlo simulations were used to estimate the absolute magnitude of \epic in the $K_s-$band, $M_{K_s}$ from the stellar radius and metallicity. The empirical relations for M-dwarfs of \cite{Mann15} were used to determine the stellar mass, \mstar, from $M_{K_s}$ and \feh. The distance to \epic was determined using $M_{K_s}$ and the apparent $K$-band (2MASS) magnitude, and is also reported in Table~\ref{tab:spec}.

\begin{table}
\caption{Stellar parameters from analysis of the Keck/HIRES spectra}
\begin{center}
\begin{tabular}{ll} \hline
Parameter  &  Value \\ \hline
\teff /K &  $3492 \pm 70$\\
\rstar / \rsol &  $0.442 \pm 0.044$\\
\feh (dex) &  $0.08 \pm 0.12$\\
\mstar / \msol  & $0.463 \pm 0.052$\\
\logg [cgs] & $4.815 \pm 0.043$\\
\lstar / \lsol & $0.0264 \pm 0.0058$\\
Distance / pc  & $95 \pm 14$\\
Spectral type & M3\,V\\
\hline
\end{tabular}
\end{center}
\label{tab:spec}
\end{table}

\section{Excluding non-planetary scenarios}
\label{sec:scenarios}

Because of the small stellar radial velocity amplitude expected for a planet of this size (Section~\ref{sec:rv:expected}), and the difficulty in achieving high-precision radial velocities for M-dwarfs, we must confirm the planetary nature of the system in other ways. We consider each of the possible non-planetary explanations for the system, and use data from a variety of sources to argue against each of them.

The most plausible alternative explanations (besides a transiting planet) for the \ktwo photometry of \epic are:
\begin{itemize}
\item The eclipse signal is from a background binary system, which is blended with the target star (the BEB scenario).
\item The system is a hierarchical triple, composed of the M-dwarf and two fainter, eclipsing companions.
\item The observed eclipses are caused by a white dwarf in a mutual orbit with the target.
\item The system is an M-dwarf -- M-dwarf binary, with approximately equal-depth primary and secondary eclipses.
\end{itemize}

In the following sections, we place constraints on each of these scenarios.

\section{Limits on background objects}
\label{sec:bg}
The two closest objects observed by \ktwo (EPIC~228814238 to the north east, and EPIC~228813721 to the south east of the target; Fig.~\ref{fig:archive}) show no variability when phased with the ephemeris of \epic. This demonstrates that the signal is not due to stray light from another \ktwo target. There are no known sources within 20\arcsec of the target in any catalogue in Vizier \citep{Vizier}.

\subsection{Archival images}
\label{sec:archive}

\begin{figure*}
	\includegraphics[width=5.5cm]{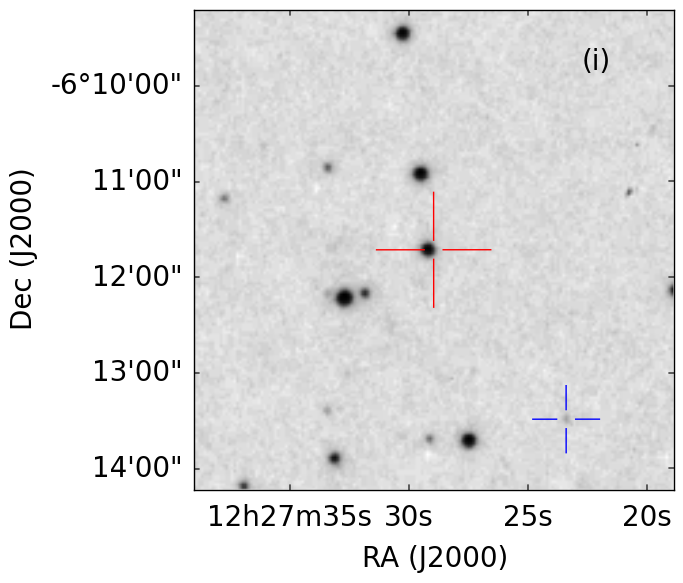}
    \includegraphics[width=5.5cm]{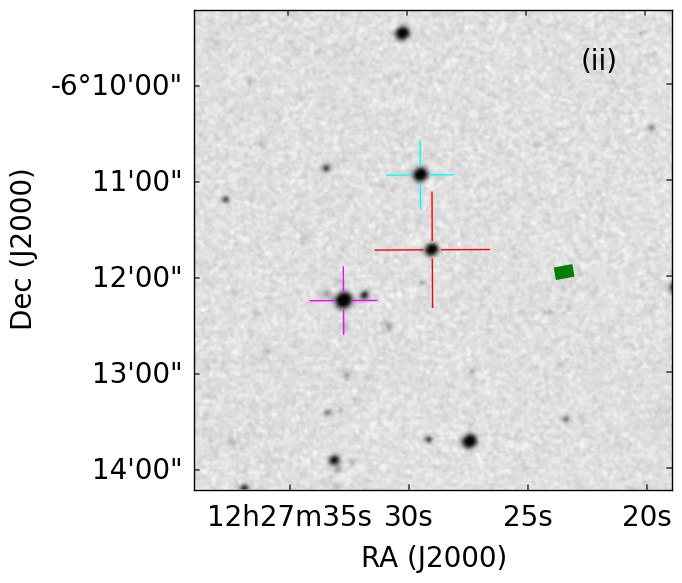}
    \includegraphics[width=5.5cm]{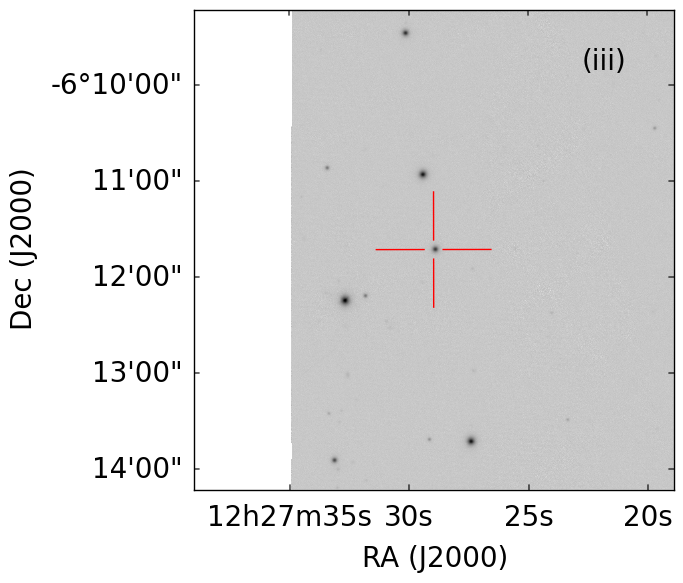}
    \caption{Archival images of \epic, demonstrating its proper motion over nearly six decades. The images are from (i) 1954, (ii) 1992, and (iii) 2012. Each image is 5\arcmin ~by 5\arcmin, and in each case North is up and East is to the left. The position of \epic (J2000 epoch) is indicated with a red reticle. The blue reticle in the leftmost panel indicates the position of the star used to determine the limiting magnitude of the image (see Section~\ref{sec:archive} for full details). The arbitrarily-positioned green rectangle in image (ii) indicates the size of the photometric aperture used by \citet{vburg} to extract the flux of \epic. The cyan (NE of target) and magenta (SE of target) reticles in image (ii) indicate the positions of EPIC~228814238 and EPIC~228813721, respectively. The provenance and dates of observation for these images are given in Table~\ref{tab:archive}.}
    \label{fig:archive}
\end{figure*}

We searched various archives for images of \epic, in order to see if, given its relatively large proper motion ($\approx 80 ~\mathrm{mas yr^{-1}}$) it is possible to see that the star has moved sufficiently that we can rule out the presence of an unbound background contaminant. In order to have a chance of seeing this, we need images spanning several decades. Using the STScI Digitized Sky Survey\footnote{http://stdatu.stsci.edu/cgi-bin/dss\_form}, we found an image dating back to 1954 (Fig.~\ref{fig:archive}, Table~\ref{tab:archive}). Using the Digital Access to a Sky Century @ Harvard (DASCH)\footnote{http://dasch.rc.fas.harvard.edu/project.php} we found several significantly older plates covering our target. Unfortunately, however, almost all of the plates older than 1954 do not go sufficiently deep to show a star as faint (particularly at the blue wavelengths to which photographic plates are most sensitive) as \epic. Our target is visible in one plate from 1920, but it is clearly close to the detection threshold; its value is limited since we would not be able to see any fainter background objects that may exist.

From the three images shown in Fig.~\ref{fig:archive}, we can see that as expected \epic has moved several arcseconds since the first available epoch (1954). By comparing that image to the latest epoch (2012), we can determine that there is no background object coincident with its current position visible in the 1954 plate. In order to estimate the limiting magnitude of the 1954 image, we consider an object nearby to our target, approximately 140\arcsec  to the southwest. This object, whose position is indicated by a blue reticle in the leftmost panel of Fig.~\ref{fig:archive}, is clearly above the detection threshold of the image. We identify this object as 2MASS~J12272341-0613291, which has $J = 15.473\pm0.056$ and a $J-K$ colour of $0.91\pm0.11$ (\epic has $J = 11.764 \pm 0.026$ and $J-K = 0.882\pm0.035$). We therefore conclude that the 1954 plate is sensitive to objects about 4 magnitudes fainter than \epic, or down to a \kep magnitude of about 18.5.

\begin{table}
\caption{Details of the archival images shown in Fig.~\ref{fig:archive}. The indices in the first column correspond to those in the upper right corner of each panel of Fig.~\ref{fig:archive}}
\begin{tabular}{llll} \hline
Image  & Source & Date & Filter \\ \hline
(i) & POSS-I$^\dagger$ & 1954 May 24 & `red' \\
(ii) & POSS-II & 1992 March 01 & `red'\\
(iii) & Pan-STARRS$^*$ & 2012 January 28 & $g$\\
\hline
\end{tabular}
\label{tab:archive}\\
$^\dagger$ Palomar Observatory Sky Survey\\
$^*$ Panoramic Survey Telescope and Rapid Response System
\end{table}

\subsection{AO imaging}
\label{sec:ao}

\begin{figure}
    \includegraphics[width=\columnwidth]{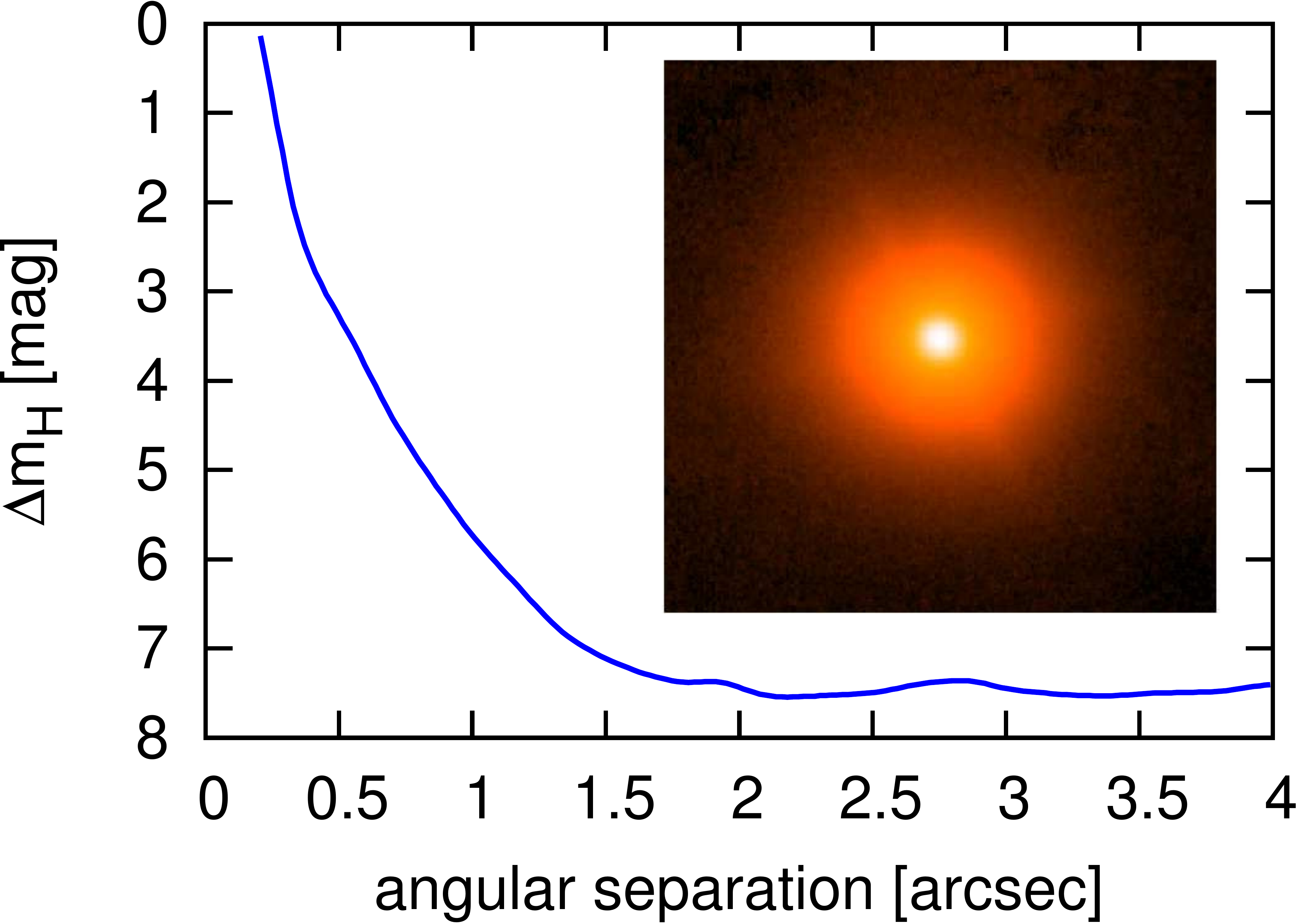}
    \caption{Contrast curve calculated from the saturated Subaru/IRCS adaptive optics image. The curve is a 5-$\sigma$ upper limit to the difference in $H$-band magnitudes between the target and a putative neighbouring object, as a function of angular separation. Inset: the 4\arcsec $\times$ 4\arcsec  ~Subaru/IRCS image.}
    \label{fig:ao}
\end{figure}

The AO imaging data was reduced following the procedure described in \cite{Hirano16}; we applied dark-subtraction, flat-fielding, distortion-correction, before aligning and median-combining the saturated and unsaturated frames separately. The full width at half maximum (FWHM) was measured for the unsaturated combined image to be $\sim 0.17\arcsec$. We then convolved the combined saturated image adopting a convolution radius of FWHM/2 and computed the standard deviation of the flux counts in each annulus centered on \epic. Figure \ref{fig:ao} shows the resultant $5-\sigma$ contrast curve as a function of angular separation from \epic's centroid, along with the AO image itself. We achieved $5-\sigma$ contrasts of $\Delta m_H=5.8$ mag and 7.4 mag at 1\arcsec and 2\arcsec. This acts to rule out the presence of objects fainter than $H=18.5$ further than 1.5\arcsec from \epic.

If the transit signal is caused by a blended background object, the brightest it can be is a \kep magnitude of about 18.5. In that case, such an object would have to undergo a $\sim 1$ per cent eclipse to account for the observed eclipse depth of approximately 350~ppm. The faintest such an object could be is if it underwent a total eclipse and had a magnitude of around 23 in the \kep bandpass.

~\\
The area of the photometric aperture employed by \cite{vburg} for \epic is $2\times3$ pixels, or 95\arcsec$^2$ (the \kep pixel scale is 3.98\arcsec~pixel$^{-1}$). By simulating a 1~sq.deg. area of sky centered on \epic using the {\sc TRILEGAL} Galactic model \citep{trilegal}, we estimate that there is a 4~per~cent chance of a star with a \kep magnitude of between 18.5 and 23.0 falling inside the photometric aperture. In Section~\ref{sec:dis:scen}, we modify this probability by considering the chance that the star is a binary system of a type that could produce the observed transit signal.

\subsection{Speckle imaging}

\begin{figure}
	\includegraphics[width=\columnwidth]{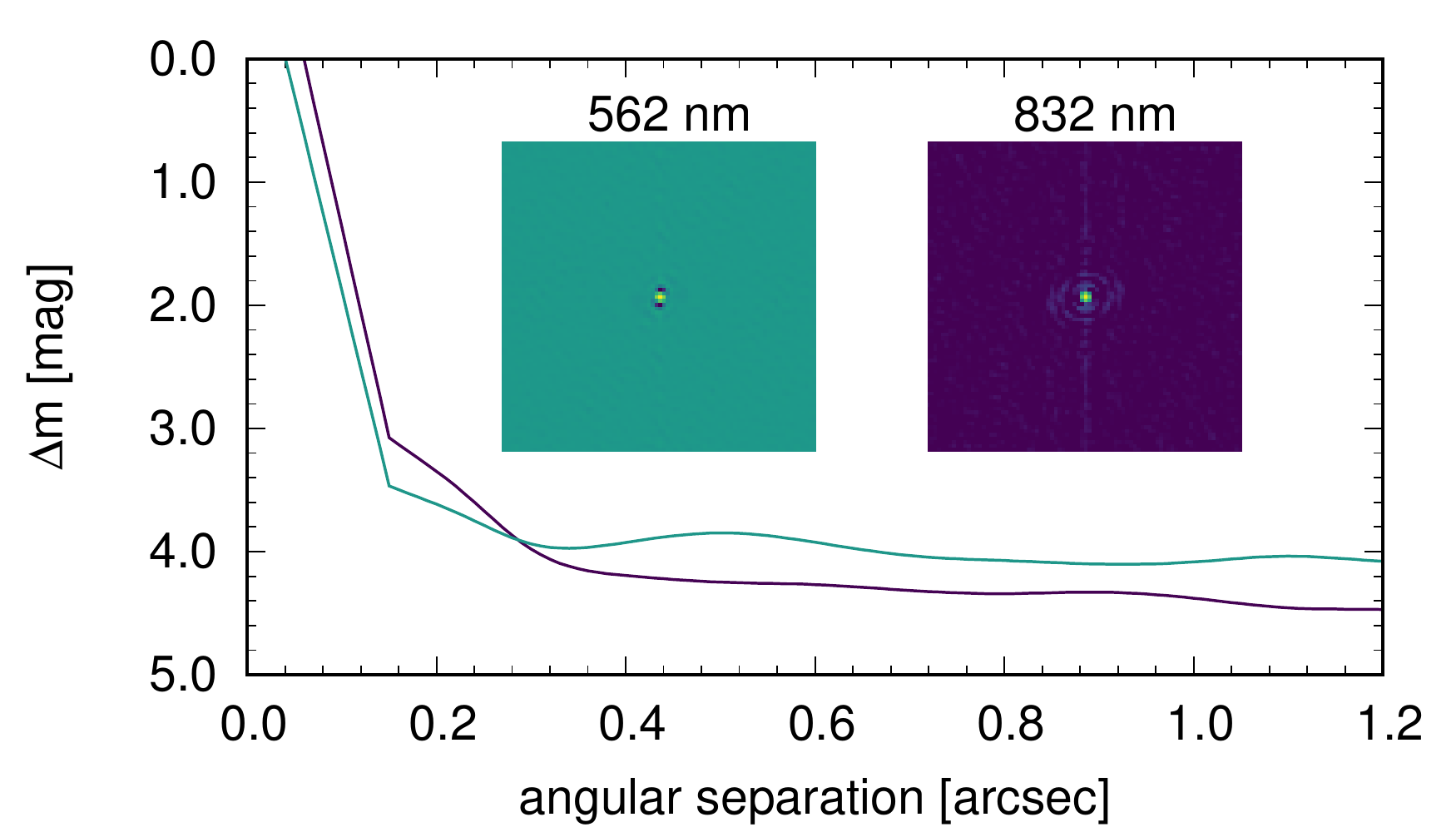}
    \caption{Contrast curves calculated from our WIYN/NESSI speckle images (inset). The green curve corresponds to the 562~nm image, and the purple curve to the 832~nm image. Each of the reconstructed images is $\approx 1.2\arcsec \times 1.2\arcsec$ in size.}
    \label{fig:speckle}
\end{figure}

Using the point source calibrator images, we computed reconstructed $256 \times 256$ pixel images in each band, following the approach of \cite{speckle}. No secondary sources were detected in the reconstructed images, which are $\sim 4.6\arcsec \times 4.6\arcsec$, although only the central portions are shown in Fig.~\ref{fig:speckle}. We measured the background sensitivity of the reconstructed images using a series of concentric annuli centered on the target star, resulting in 5-$\sigma$ sensitivity limits (in $\Delta$-mags) as a function of angular separation.

The 832~nm contrast curve and reconstructed image are shown in Fig.~\ref{fig:speckle}, which indicates that objects up to about 4~mags fainter than \epic are excluded within about 0.25\arcsec. This acts as further confirmation of the results we obtained from the archival images in Section~\ref{sec:archive}.

\section{Radial velocity constraints}

\subsection{Expected amplitude of RV signal}
\label{sec:rv:expected}
If the secondary is indeed a planet, we can use the planetary-to-stellar radius ratio we measure from modelling the light curve, along with our stellar radius from the spectral analysis to calculate the planetary radius. By comparing this planetary radius to that of other known exoplanets with mass measurements, we can estimate the planetary mass we might expect.

For a stellar radius of 0.44~\rsol and a radius ratio of $\approx 0.02$, we derive a planetary radius of around 1.0~\re. For such a planet, we consider a very conservative upper-limit to the mass to be 3~\me (\citealt{Seager07} predict a mass of around 2.7~\me for a 1~\re planet composed of solid iron).
The radial velocity semi-amplitude, $K$, is given by,
\begin{equation}
K = \left(\frac{2\pi G}{P}\right)^{\frac{1}{3}} \frac{M_\mathrm{p}\sin i}{\left(M_* + M_\mathrm{p}\right)^{\frac{2}{3}}\sqrt[]{1-e^2}}
\end{equation}
Assuming $i=90\deg$, $e=0$, and that $M_* >> M_{\mathrm{p}}$, we find $K < 6$~\ms. For a stellar-mass object, this velocity would be four or more orders of magnitude larger.

\subsection{Upper limit to the planet mass from RVs}
\label{sec:rv}

Using our NOT/FIES and Keck/HIRES radial velocity measurements (Section~\ref{sec:keck}, Table~\ref{tab:rv}), we can place an upper limit on the mass of an orbiting companion to \epic. Using the orbital period and epoch of transit determined from the \ktwo photometry (Section~\ref{sec:tlcm}), we calculate the orbital phase of the RV data. Given that the period and phase of the RV curve are determined from the photometry, and assuming a circular orbit (Section~\ref{sec:circ}) the only unknowns are the systemic radial velocity, $\gamma$, and the orbital velocity semi-amplitude, $K$. We determine $\gamma$ and the 3-$\sigma$ upper limit to $K$ by means of a simple $\chi^2$-minimisation routine, and assume that there is no instrumental offset between FIES and HIRES. We account for the lengthy RV exposure times (particularly in the case of the FIES data) by sampling the model at five minute intervals, and using numerical integration to determine the effective model value at each RV datum. We note that the resulting 3-$\sigma$ upper limit to $K$ is similar to the values obtained by considering only the FIES or HIRES data alone.

\begin{figure}
	\includegraphics[width=\columnwidth]{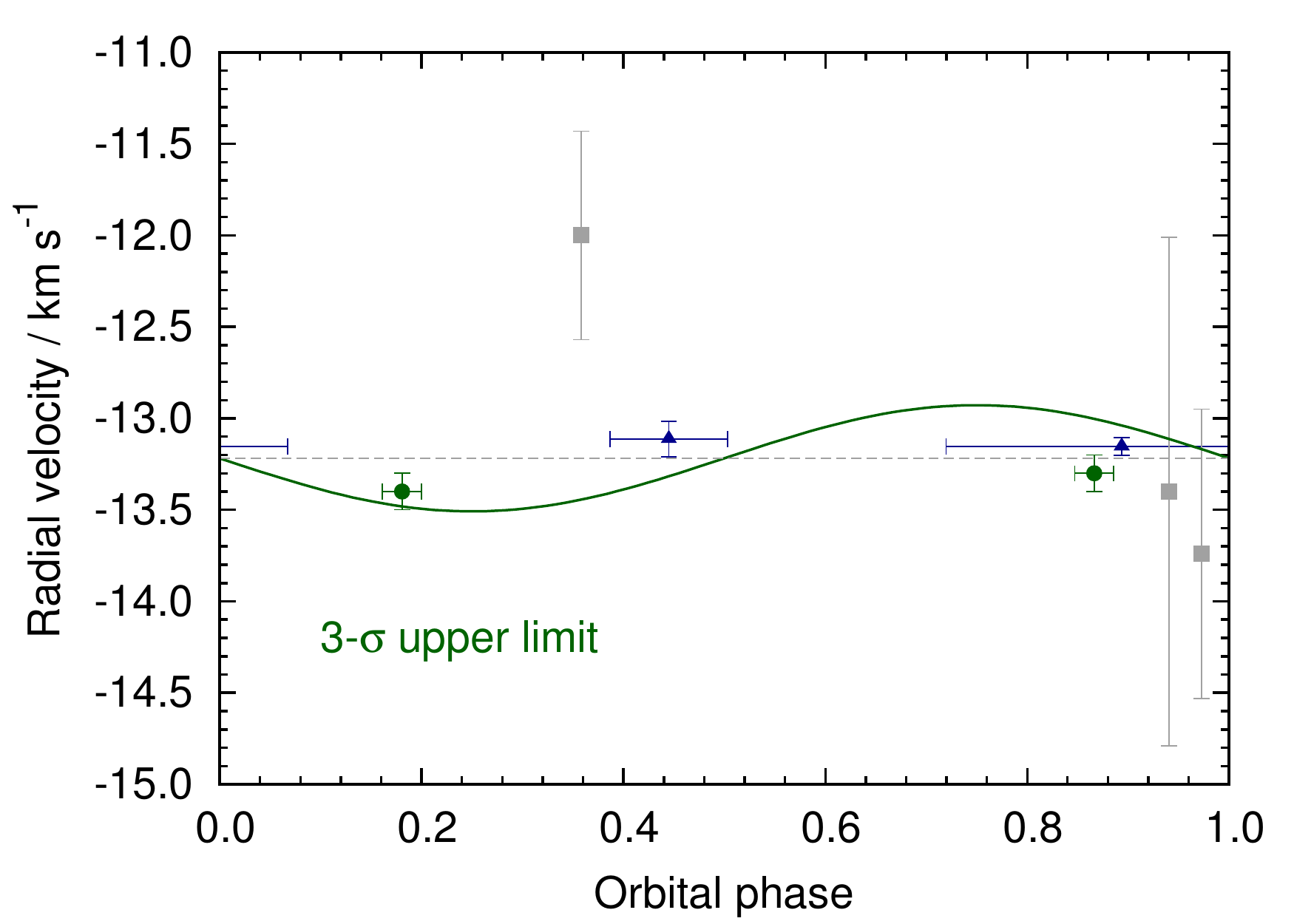}
    \caption{Keck/HIRES (green circles) and NOT/FIES (blue triangles) radial velocities and 1-$\sigma$ error bars, over-plotted with a model (solid line) with $K$ = 290 \ms, corresponding to the 3-$\sigma$ upper limit to $K$. Horizontal error bars are used to indicate the exposure times of the Keck and FIES data. The Subaru/IRCS radial velocities are shown as grey squares, but are not used to calculate the upper limit to $K$. The dashed line represents the systemic radial velocity, $\gamma$.}
    \label{fig:rv}
\end{figure}

The resulting 3-$\sigma$ upper limit to $K$ is 290~\ms (Fig.~\ref{fig:rv}), corresponding to a planet mass of around 0.5~\mjup. We can therefore conclude that if the 4.3-hr photometric variation is caused by a body orbiting the target star, it must have a mass firmly within the planetary regime. Repeating the above procedure, but phasing the RV data on twice the orbital period, gives 3-$\sigma$ upper limits of 145~\ms on $K$, and 0.3~\mjup on \mplanet.

We also extracted RVs from our three Subaru/IRCS spectra. The data were reduced using the standard IRAF procedure: bias-subtraction, flat-fielding, cosmic-ray removal, before extracting 1-dimensional (1D) spectra (including sky subtraction). The wavelength was calibrated using the OH emission lines for each frame. To correct for the telluric absorption lines, the 1D spectrum of \epic was divided by the normalised spectrum of HIP~61318 for each set (1 set = 1 ABBA nodding). The RV was then estimated by computing the cross-correlation between the observed spectra and the PHOENIX model template (Allard et al. 2011) mimicking \epic's spectrum ($T_\mathrm{eff}=3500$ K). We also computed the cross-correlation between the model telluric transmission spectrum and HIP~61318's spectrum for each set to enhance the accuracy of the wavelength calibration. The final RV values were calculated by inspecting the peaks of those cross-correlation functions. 

These RVs are plotted in Fig~\ref{fig:rv}, alongside those from NOT/FIES and Keck/HIRES. All three datasets are consistent with each other within $\sim 2\sigma$, although the precision of the Subaru RVs is much worse because of IRCS's lower spectral resolution and poor pixel sampling.

\section{Light curve modelling}
\label{sec:tlcm}

\subsection{Basic fit}
\label{sec:basic_fit}
\subsubsection{The TLCM code}

We model the \ktwo light curve of \epic as reduced by \cite{vburg} using the {\sc Transit and Light Curve Modeller (TLCM)} code. {\sc TLCM} has been used to model exoplanet light curves and radial velocities in numerous previous studies, including planets discovered in long-cadence \ktwo data \citep[e.g. K2-99b;][]{K299}. The code is described in \cite{Szilard_BD}, and a more detailed description will accompany the first public release of the code (Csizmadia in prep.).

In brief, TLCM employs the \cite{M&A} model to fit the photometric transit, compensating for \ktwo's long exposure times using numerical integration. The code uses the combination of a genetic algorithm followed by simulated annealing. The former is used to find the approximate global minimum, and the latter to refine the solution, and explore the neighbouring parameter space for the determination of uncertainties on the model parameters. TLCM allows the user to include emission from the secondary object (planet), the beaming effect, ellipsoidal variability, and the reflection effect in the photometric model. TLCM is also capable of incorporating radial velocity data (including that covering the Rossiter-McLaughlin effect) from multiple instruments into its global fit.

\subsubsection{Fitted parameters}

For our basic fit, we fit for the following parameters: the orbital period, $P$, the epoch of mid-transit, $T_0$, the scaled semi-major axis ($a/\rstar$), planet-to-stellar radius ratio ($\rplanet / \rstar$), and the impact parameter, $b$.

\subsubsection{Limb darkening}
We initially fitted for the limb-darkening coefficients, but found them to be poorly-constrained by the data. Instead, we decided to adopt values from \cite{Sing09} appropriate for \epic and \ktwo. Interpolating the tabulated coefficients for $[M/H] = 0.08$ and \logg = 4.8 at \teff = 3500~K gives $u_1 = 0.427$ and $u_2 = 0.317$. Then, $u_+ = u_1 + u_2 = 0.744$ and $u_- = u_1 - u_2 = 0.110$. We find that fixing the limb-darkening coefficients in this way has a negligible impact on the results of the fit.

\subsubsection{Circular orbit}
\label{sec:circ}
Since we do not have RV data which would allow us to determine the eccentricity of the planet's orbit, we impose a circular-orbit solution. At such close orbital distances, both the theoretical and the empirical expectation is that the planet's orbit will be circular. Using Equation~1 of \cite{Jackson08}, we estimate a tidal-circularisation time-scale of just 2~Myr. We do not have a formal age estimate of the system, but suggest that \epic is extremely unlikely to be just a few Myr old. Young M-dwarfs are likely to show significant flaring and stellar activity, which we do not observe, and in any case it is unlikely that we happen to observe such a long-lived star in the first fraction of a per cent of its lifetime. Finally, the ALFOSC and HIRES spectra shows no Li\,{\sc i} $\lambda$6708~{\AA} absorption line. Since lithium is rapidly depleted in the convective layers of low-mass stars in the early phases of stellar evolution, we conclude that \epic cannot be a pre-main sequence star.

\subsection{Further light curve tests}

\subsubsection{Odd / even transits}
\label{sec:odd-even}
\begin{figure}
	\includegraphics[width=\columnwidth]{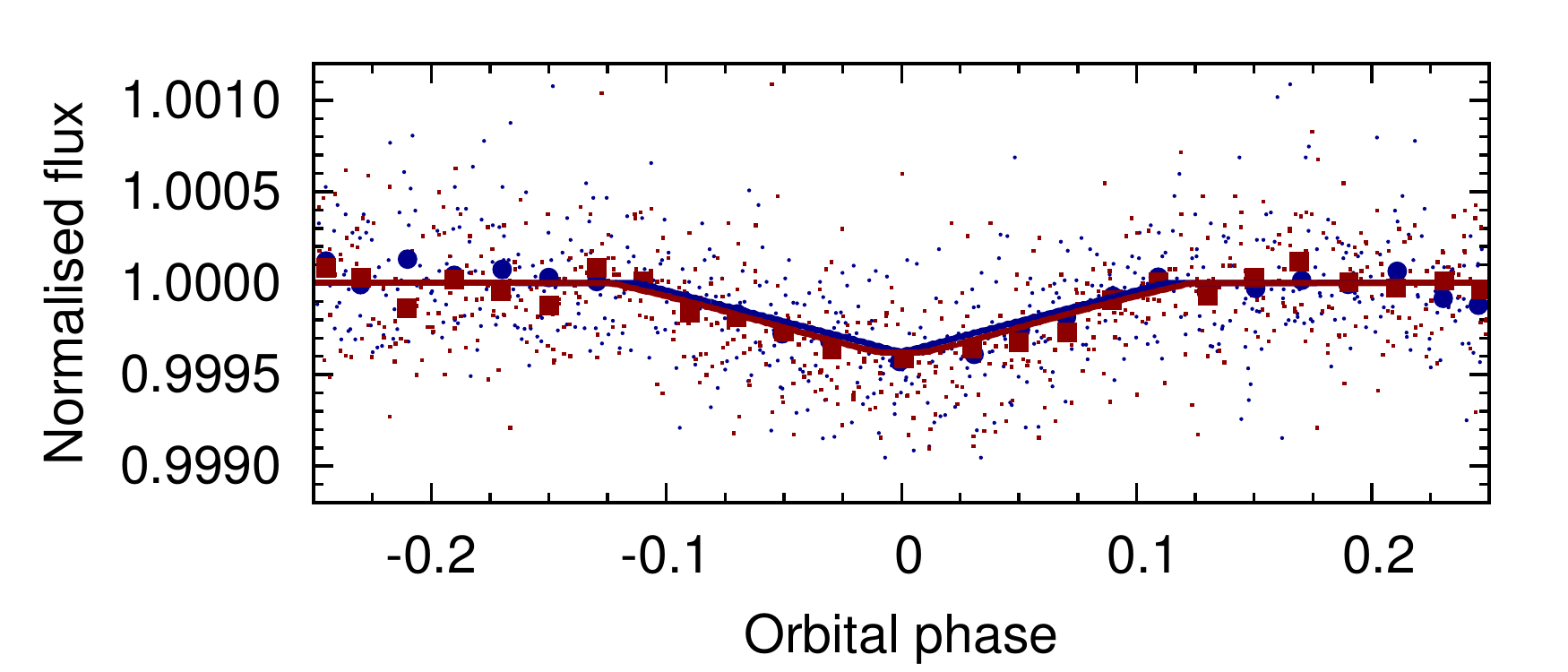}
    \caption{\ktwo photometry, for odd-numbered transits (blue circles) and even-numbered transits (red squares), over-plotted with binned data (larger symbols). Our best-fitting transit models are also shown, in colours corresponding to the datapoints. No significant difference between odd and even-numbered transits is seen (see Section~\ref{sec:odd-even}).}
    \label{fig:odd-even}
\end{figure}

We repeated the TLCM modelling with modified light curves to include only the odd-numbered transits in one run, and only the even-numbered transits in another. For this modelling, we selected data points only in the phase range $-0.25 < \phi < 0.25$. The data cover a total of 151 odd-numbered transits, and 152 even-numbered transits. We found planet-to-stellar radius ratios consistent with each other to within 1-$\sigma$ (Fig.~\ref{fig:odd-even}). All other fitted parameters were similarly consistent. We find a ratio between the odd and even depths of $\delta_\mathrm{even} / \delta_\mathrm{odd} = 1.07 \pm 0.10$ ($\delta_\mathrm{even} - \delta_\mathrm{odd} = 25 \pm 35$~ppm). We therefore conclude that there is no evidence to an odd/even transit depth difference, which would be indicative of a binary star system with unequally-sized components, and an $\approx 8.6$-hour orbital period.

\subsubsection{Occultation depth}

Using our best-fitting planetary parameters, we calculated the equilibrium temperature of the planet, assuming zero albedo (Table~\ref{tab:tlcm}). Using this temperature, an appropriate ATLAS9 synthetic spectrum ($\teff = 3500$~K, $\feh = 0$, $\logg = 5.0$; \citealt{Castelli2004}), and the Kepler response function, we calculated the expected occultation depth from thermal emission alone. The predicted depth is just 0.15~ppm, increasing to 2.5~ppm if we use the higher equilibrium temperature we derive for the immediate re-radiation case. We note that \cite{kepler78} detect a $10.5 \pm 1.2$~ppm occultation for Kepler-78b, which they attribute to a combination of thermal emission and reflected light. \cite{KOI1843}, however, detect no occultation of KOI~1843.03, but place a 3-$\sigma$ upper-limit of 18~ppm on the occultation depth from the \kep photometry. The planet-to-star area ratio of \epic is a factor of $\approx 1.7$ larger than for Kepler-78, so for a given albedo one would expect a greater contribution to the occultation depth from reflected light in this case.

We performed a fit of the light curve where we also fitted for an occultation centred on phase 0.5. We find a best-fitting occultation depth of $4 \pm 10$~ppm and hence conclude that the 3-$\sigma$ upper-limit to the occultation depth is 33~ppm, consistent with the planetary scenario described above.

\subsubsection{Out-of-eclipse variation}
\label{sec:oot}

We also fitted for out-of-eclipse variation, specifically that caused by the beaming effect, and by ellipsoidal effects. Fitting for the beaming effect gives a best-fitting radial velocity semi-amplitude, $K = -10 \pm 228$~\ms. Combining this with the stellar mass (Table~\ref{tab:spec}) results in a 3-$\sigma$ upper limit to the planet mass of 356~\me (1.1~\mjup). The best-fitting ellipsoidal amplitude is $3 \pm 17$~ppm, giving 2-$\sigma$ and 3-$\sigma$ upper limits of 34~ppm and 55~ppm, respectively.

\subsubsection{Fitting with twice the orbital period ($\approx$ 8.6~ h)}

In order to consider the BEB scenario (Section~\ref{sec:scenarios}, \ref{sec:dis:scen}), we model the light curve and force the orbital period to be approximately twice that of the planetary orbital period. We also fit for out-of-eclipse variation (as in Section~\ref{sec:oot}). We ran three instances of the model fitting: one where we fitted for the contamination from third light, one where this contamination was set to zero, and one where it was fixed to the maximum allowed value of $l_3 = f_\mathrm{contamination} / f_\mathrm{EB} = 2500$ (from conversion of the magnitude limits placed in Section~\ref{sec:bg}).

When third light was not allowed, {\sc TLCM} did not converge on a good fit to the data. The other cases produced similar-looking fits, with the best-fitting $l_3 = 1616 \pm 212$. The best-fitting amplitude of the ellipsoidal variations from this case was $18.5 \pm 8.5$~ppm, leading to 2-$\sigma$ and 3-$\sigma$ upper limits of 35 and 44~ppm, respectively.

\subsection{Results}
\label{sec:results}

\begin{figure*}
	\includegraphics[width=\textwidth]{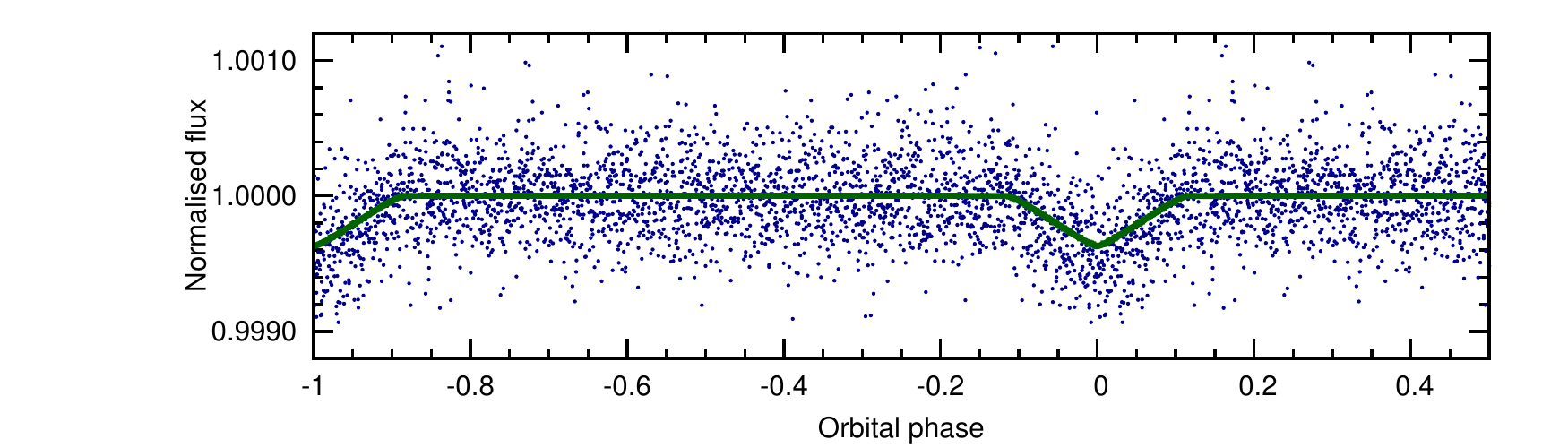}
    \caption{\ktwo photometry, over-plotted with our best-fitting transit model, as described in \ref{sec:results}. Note that the apparent 'V'-shape of the transit is caused by the cadence of the data (see Section~\ref{sec:texp} and Fig.~\ref{fig:texp}).}
    \label{fig:lc}
\end{figure*}

The parameter values resulting from our fit to the photometry (as described in Section~\ref{sec:basic_fit}) are presented in Table~\ref{tab:tlcm} and the light curve fit is shown in Fig.~\ref{fig:lc}. Our best-fitting stellar density from the transit observations is $(13.0 \pm 2.0) \times 10^3$~kg~m$^{-3}$, which we can compare to the stellar density we derive from spectra in Section~\ref{sec:star}: $(7.6 \pm 2.4) \times 10^3$~kg~m$^{-3}$. These two values are compatible at a level  $<$~2-$\sigma$, which provides further support for the planetary hypothesis. 

\begin{table*} 
\caption{System parameters from {\tt TLCM} modelling} 
\label{tab:tlcm}
\begin{tabular}{lccc}
\hline
\hline
Parameter & Symbol & Unit & Value\\
\hline 
\textit{{\tt TLCM} fitted parameters:} &\\
&\\
Orbital period	    	    	    	    & $P$ & d & $0.179715 \pm 0.000001$\\
Epoch of mid-transit	    	    	    & $T_{\rm 0}$ &$\mathrm{BJD_{TDB}}$ & $ 2457583.0242 \pm 0.0004 $\\
Scaled orbital major semi-axis              & $a/R_{\rm *}$ &...& {}$ 2.82 \pm 0.14${}\\
Ratio of planetary to stellar radii         & \rplanet / \rstar &...& $0.0184 \pm 0.0006$ \\
Transit impact parameter                    & $b$ &...& $0.02\pm0.10$\\
Limb-darkening parameters                   & $u_+$ &...&$ 0.744$ (fixed) \\
						                    & $u_-$ &...&$ 0.110$ (fixed) \\
&\\
\textit{Derived parameters:}\\
&\\
Orbital eccentricity	    	    	    & 	$e$ &...& $0$ (adopted) \\
Stellar density     	    	    	    & 	 \densstar &$10^3$ kg m$^{-3}$ & $ 13\pm2  $\\
Planet mass 	    	    	    	    & 	\mplanet &\mjup & 0.5 (3-$\sigma$ RV upper limit)\\
Planet radius	    	    	    	    & 	\rplanet &\re & $ 0.89 \pm 0.09 $\\
Orbital major semi-axis     	    	    & 	$a$ &AU  & $ 0.0058 \pm 0.0006  $\\
Orbital inclination angle   	    	    & 	$i_\mathrm{p}$ &$^\circ$  & $ 89.6 \pm 3.3$\\
Transit duration                            & $T_{\rm 14}$ &d&$0.0262 \pm 0.0011$\\
Planet equilibrium temperature$^\dagger$ & $ T_{\mathrm{p, eql,}A = 0}$ & K & $1471 \pm 47$\\
\hline
\multicolumn{4}{l}{$^\dagger$The equilibrium temperature is calculated assuming a planetary albedo of zero, and}\\
\multicolumn{4}{l}{isotropic re-radiation. For efficient re-radiation to the nightside, the equilibrium}\\
\multicolumn{4}{l}{temperature is $1749 \pm 56$~K, and for immediate re-radiation it is $1880 \pm 60$~K.}
\end{tabular} \\ 
\end{table*} 

\subsection{The impact of \ktwo's long exposure time}
\label{sec:texp}
\begin{figure}
	\includegraphics[width=\columnwidth]{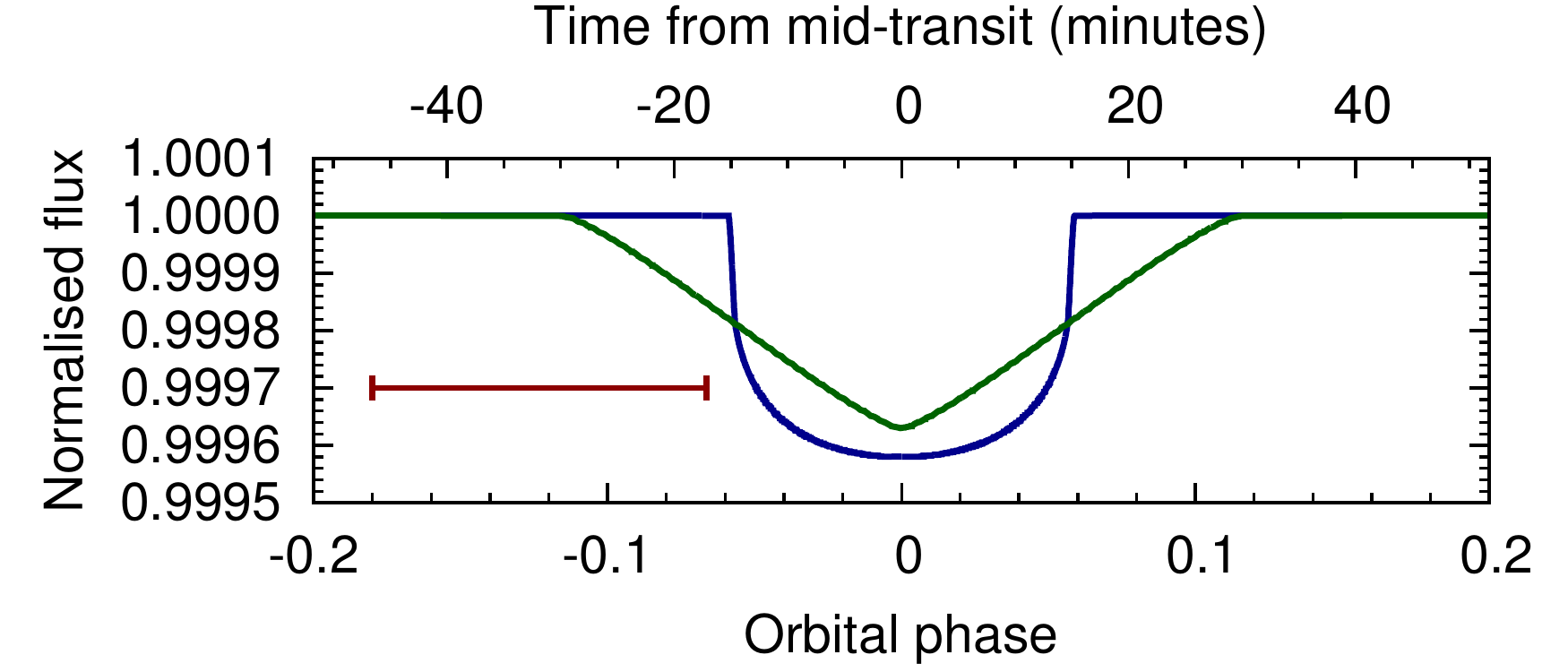}
    \caption{Comparison of our best-fitting model (green curve) with the same model if the exposure time were short (1~s, blue curve). The long-cadence \ktwo data has the effect of making the transit longer, shallower, and more 'V'-shaped. The red bar indicates the duration of a single \ktwo exposure.}
    \label{fig:texp}
\end{figure}

As has been shown previously from both theoretical \citep{Kipping10} and observational \citep[e.g.][]{Muirhead13} standpoints, the effect of \ktwo's relatively long exposure time (c. 30 minutes) is to `smear out' and make more 'V'-shaped the transit in the light curve. Because of the extremely short-period orbit of \epic~b, and the fact that the transit duration is comparable to the exposure time, the effect is exaggerated in this case. To illustrate this effect, in Fig.~\ref{fig:texp} we compare the transit shape that would be observed for a short (1~s cadence) exposure time with that observed for the sampling rate of \ktwo.

\section{Discussion}
\subsection{Further arguments against non-planetary scenarios}
\label{sec:dis:scen}
\subsubsection{Eclipsing binary system}

In Section~\ref{sec:scenarios}, we laid out a series of non-planetary scenarios which could explain the photometric eclipse signal observed in the light curve of \epic. The latter two of these possibilities, which involve a stellar-mass object in orbit around an M-dwarf are excluded by our RV measurements (Section~\ref{sec:rv}).

The M-dwarf -- M-dwarf binary possibility can also be excluded from the photometry alone. Here, the two components of the binary would be near-identical, since we detect no difference between odd and even-numbered transits (Section~\ref{sec:odd-even}). In this scenario, the orbital period is twice that if the system is planetary, at around 0.36~d. By comparing our phased light curve to those presented in \cite{Matijevic12} (their Fig.~3), we determine that the morphology parameter of our light curve is certainly less than 0.4, and probably significantly smaller than that. By comparing this period and morphology parameter to those of an ensemble of known EBs, we conclude that \epic cannot be an EB.

\begin{figure}
	\includegraphics[width=\columnwidth]{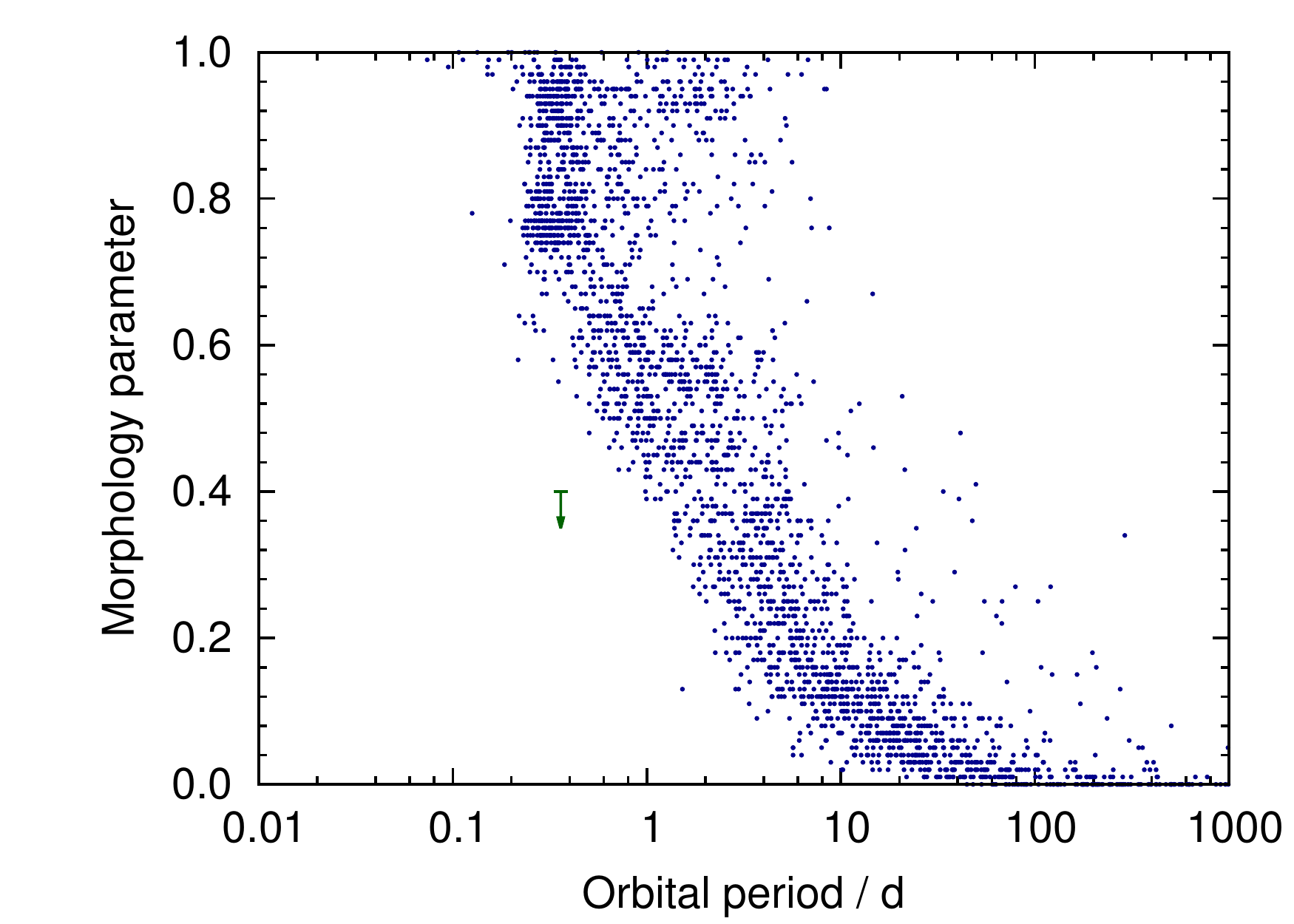}
    \caption{Light curve morphology parameter, $c$, as a function of orbital period for the EB systems in the Kepler Eclipsing Binary Catalog. \epic is shown with an arrow indicating our conservative upper limit on $c$, at twice the orbital period of the planetary system. There are no EBs in the region occupied by \epic.}
    \label{fig:KEBC}
\end{figure}

Of the 2876 EBs in the Kepler Eclipsing Binary Catalog\footnote{http://keplerebs.villanova.edu/} \citep[KEBC][]{Kep_binary_cat}, 1\,403 have a morphology parameter, $c$, less than or equal to 0.4. The minimum orbital period of these objects is 0.98~d. In other words, there exist no EBs in the period range of \epic that are not contact or semi-detached systems. Filtering the KEBC by period instead, we find that there are 355 objects with $P \leq 0.36$~d, but none of these has $c < 0.55$. This is illustrated in Fig.~\ref{fig:KEBC}, where we plot $c$ against $P$ for the objects in KEBC.

\subsubsection{Background eclipsing binary system}

One non-planetary explanation that remains is that of a blended background eclipsing binary (BEB). In the most-likely form of this scenario, the target star is blended with a distant EB, whose components are of approximately equal radius and mass. The total flux from such a system is constrained to be between about $1/40$ and $1/2500$ of the flux from \epic (from conversion of the magnitude limits placed in Section~\ref{sec:bg}).

The amplitude of the ellipsoidal variability of the BEB will be diluted by light from the much-brighter target. Our 2-$\sigma$ upper limit of 35~ppm on the amplitude of the ellipsoidal variability of the light curve then corresponds to a less-stringent limit on the un-diluted amplitude of variation for the BEB. For the faintest possible case, with the strongest dilution, our $2-\sigma$ limit is 8.5 per cent.

To calculate the expected ellipsoidal variability amplitude of such a system, we use the relation of \cite[][their Equation 1]{Morris93}. We assume a mass ratio, $q$, of the binary components of unity, an inclination angle of 90\degr, and conservative values of 0.9 and 0.0 for the limb-darkening and gravity-darkening coefficients, $u_1$ and $\tau_1$, respectively. We conservatively assume the ratio of the primary stellar radius to the semi-major axis ($R_1/A$ in the notation of \citealt{Morris93}) to be 0.1 (which implies a stellar mass of 0.15~\msol for each component). Larger, more plausible values would result in an even larger ellipsoidal amplitude. We calculate an expected ellipsoidal amplitude of 8~per cent.

We can therefore conclude that almost all plausible binary star configurations would result in an observable ellipsoidal variation in the \ktwo light curve, which is not detected.

In Section~\ref{sec:ao} we used the {\sc TRILEGAL} Galactic model to estimate the probability that there is a faint background star in the photometric aperture. We can place further constraints with {\sc TRILEGAL} on the probability that a binary system of a type that could produce the observed light curve when blended with \epic\footnote{We use the {\sc TRILEGAL} default values for the binary fraction and minimum mass ratio, $f_b = 0.3$ and $b_b = 0.7$ \citep{trilegal}.}. Such a binary must have a mass ratio close to unity; we estimate from our 3-$\sigma$ limit on the difference between the odd and even-numbered transits, and the main-sequence relation ${\mstar}/{\msol} \approx \left({\rstar}/{\rsol}\right)^{5/4}$ that the smallest the mass ratio can be is 0.78. The smallest values of ellipsoidal modulation, which are marginally permitted by our upper limit are those arising from M-dwarf binaries. We can thus also limit the colour of the binary systems, to those with $J-K > 0.85$. The probability of such a binary with these properties, and of the right brightness, falling within the photometric aperture is $3.4 \times 10^{-5}$, or about 1-in-30\,000. We can further reduce this probability (by a factor of a few), since we require an eclipsing alignment for the binary. Multiplying the occurrence rate of \cite{Sanchis-Ojeda14} (see Section~\ref{sec:mdwarf}) by the geometric transit probability for \epic~b gives a probability of $3.9 \times 10^{-3}$, more than two orders of magnitude greater than the BEB probability.

\subsubsection{Bound triple star system}

A similar line of reasoning to that used to quantify the likelihood of the BEB scenario can also be applied to the hierarchical triple scenario, in which \epic is accompanied by a pair of fainter, eclipsing companions.  No meaningful limits on the brightness of these components exist from our AO or speckle imaging, since they could lie well within a single arcsecond of the target. The fainter companions could be up to 8.6 magnitudes fainter than \epic, if they completely eclipse each other. From tabulations of absolute magnitudes for late-type stars\footnote{\label{mamajek}http://www.pas.rochester.edu/$\sim$emamajek/ EEM\_dwarf\_UBVIJHK\_colors\_Teff.txt \citep{Pecaut_Mamajek}}, this encompasses all the late M-dwarfs, and perhaps also very early L-dwarfs. A pair of M7V companions is likely to result in a detectable degree of ellipsoidal variability, but a pair of M9V stars may not.

Estimating the probability that an M3 dwarf is part of a hierarchical triple with a pair of M9 stars is not straightforward, given that the full statistics of stellar multiplicity, particularly amongst late-type stars are not known. \cite{Tokovinin08} suggest that 8 per cent of solar-type stars are in triples, but the equivalent value for M-dwarfs is unknown, so we use the solar-type value. Our {\sc TRILEGAL} results suggest that 0.1 per cent of binaries have a mass ratio greater than 0.78, and a $(J-K)$ colour greater than 1.2 (an M9 star has $J-K = 1.23$$^{\ref{mamajek}}$). Multiplying these probabilities together, we find that there is an $8\times 10^{-5}$ chance of \epic having a binary companion of the correct type. Note that this excludes the requirement that the binary pair are in a very short-period orbit, and that \cite{Shan15} find that the incidence of M-dwarf binaries increases with orbital period, so the true probability that our target is a hierarchical triple is lower than this value.

Given that the calculated probabilities of a BEB or a bound triple system are significantly lower than the planetary probability, we conclude that it is overwhelmingly likely that the eclipse signal results from a terrestrial-sized planet transiting the target, \epic.

\subsection{Planet composition and mass}

Since planets must orbit outside of the Roche limit, so that stellar tidal forces do not cause them to disintegrate rapidly, we can place constraints on the composition of USP planets \citep{KOI1843}. Their Equation~5 (reproduced below) gives the minimum orbital period, $P_\mathrm{min}$ for a given mean density, $\rho_\mathrm{p}$, and central density, $\rho_\mathrm{0p}$,
\begin{equation}
P_\mathrm{min} \simeq 12.6 \mathrm{~hr} \left(\frac{\rho_\mathrm{p}}{1 \mathrm{g~cm}^{-3}}\right)^{-1/2} \left(\frac{\rho_\mathrm{0p}}{\rho_\mathrm{p}}\right)^{-0.16}
\end{equation}
Following the approach of \cite{KOI1843}, we find that $\rho_\mathrm{p} \geq 6.4~\mathrm{g~cm}^{-3}$, for $({\rho_\mathrm{0p}}/{\rho_\mathrm{p}}) \leq 2.5$. This leads to a constraint on the composition, assuming an iron core and a silicate mantle. We determine the minimum iron mass fraction to be $0.525 \pm 0.075$ (cf. 0.7 for KOI~1843.03), which is greater than that of Earth, Venus or Mars, but smaller than that of Mercury (approximately 0.38, 0.35, 0.26, and 0.68, respectively; \citealt{Reynolds69}).

The minimum density can be used along with the derived planetary radius (Table~\ref{tab:tlcm}) to calculate a lower limit to the planetary mass. We find $M_\mathrm{p,min} = 0.82 \pm 0.25$~\me. This suggests that the planet's mass lies between about 0.6 and about 2.7 times that of the Earth (see Section~\ref{sec:rv:expected} for the derivation of the upper limit).

Discovering and characterising extreme systems, such as USP planets like \epic, is important as they offer constraints for planet formation theories. Furthermore, they allow us to begin to constrain their interior structure -- and potentially that of longer-period planets too, if they are shown to be a single population of objects.

\subsection{Origins of the USP planets}

\cite{Winn_USP} recently cast doubt on the theory that USP planets are the solid cores of HJs whose gaseous envelopes have been lost. By comparing the metallicities of HJ and USP host stars, \cite{Winn_USP} concluded that unlike HJs, the USP planets do not orbit a metal-rich population of stars, and hence HJ and USP host stars constitute different populations. The remaining explanations for their existence are that the USP planets are part of the population of close-in rocky planets, or they are the remnants of smaller gas planets (mini-Neptunes). We note that \epic is not particularly metal-rich; its metallicity is only slightly above that of the Sun.

\subsection{M-dwarfs as USP planet hosts}
\label{sec:mdwarf}

\cite{Sanchis-Ojeda14} find that the USP planet occurrence rate increases with decreasing stellar mass; they calculate that $0.51 \pm 0.07$~per cent of G-dwarfs host a USP planet, compared to $1.1 \pm 0.4$~per cent of M-dwarfs. Nevertheless, it is curious that the USP planets with the shortest periods (KOI~1843.03 and \epic~b) both orbit M-dwarfs, despite there being a much larger number of G and K stars in the observed \kep and \ktwo samples. Of course, a given planet orbiting an M-dwarf will produce a deeper transit than if it orbits a star of an earlier  spectral type. \cite{Sanchis-Ojeda14}, however, determine that such USP planets in orbit around G-dwarfs are detectable in the \kep data, and \cite{Adams16} report \ktwo detections of G-dwarfs hosting USP candidates as well. Selection bias seems therefore unlikely to explain the fact that the shortest-period planets are both around M-stars.

One possibility is that short-period planets (near the Roche limit) have a longer lifetime around M-dwarfs, than around earlier-type stars. The rate of tidal decay in this regime is strongly dependent on the stellar tidal quality factor, $Q_*$, and for small eccentricities is essentially independent of the equivalent planetary quantity, $Q_\mathrm{p}$. We used the relation of \cite{Jackson08} (their Equation~2) to estimate the orbital decay timescales of \epic~b. 

Assuming a planetary mass equal to that of the Earth, we find the characteristic evolution time for \epic~b, $\tau_a = 6.3 (Q_* / 10^7) $~Gyr. The equivalent value for the same planet in an orbit of the same period around a Sun-like star is $\tau_a = 0.16 (Q_* / 10^7) $~Gyr. So, if M and G-stars have the same $Q_*$ value, the orbit of the planet around the G-dwarf should decay 40 times more quickly. This offers a tantalising hint that tidal decay rates could explain why the shortest-period planets are found around M-dwarfs. However, we caution against over-interpretation of this, given that our knowledge of stellar tidal quality factors is exceedingly limited, with estimates varying wildly over several orders of magnitude. Indeed, the theoretical study of \cite{Barker+Ogilvie} suggests that $Q_*$ may vary between $10^8$ and $10^{12}$ for F-stars alone. It would therefore seem somewhat reckless to assume that M-dwarfs and G-dwarfs should have identical $Q_*$, given the differences in their internal structures.

A final explanation for the observed phenomenon is that planet composition varies with spectral type, and that iron-rich planets such as KOI~1843.03 and \epic~b occur more frequently around M-dwarfs than other stars. Further USP planet detections are required in order to address these hypotheses.

\subsection{Future work}

It may be possible to measure the mass of \epic~b through radial velocity measurements with one of the new generation of high-resolution infrared spectrographs, such as CARMENES \citep{carmenes} or IRD \citep{IRD}. Even with such an instrument, however, the observations would be challenging, and require a significant investment of telescope time, given both the relative faintness of the target and the small amplitude of the expected RV variation. Finally, we suggest that the discovery of \epic should serve as a motivation to conduct a planet search with a low detection threshold, specifically for very-short-period systems around the M-dwarfs in the \kep and \ktwo samples.

\section{Conclusion}

Using \ktwo photometry, we have discovered a planet slightly smaller than the Earth ($\rplanet = 0.89 \pm 0.09~\re$) in an extremely short-period (4.3~hr) orbit around an M-dwarf. \epic~b joins a growing list of USP planets, and is strikingly similar to the slightly smaller KOI~1843.03 which also orbits an M-dwarf in just over 4~hr. Although we do not measure the mass of the planet, we constrain its iron mass fraction to be greater than 0.45 (1-$\sigma$). We use a combination of archival images, AO imaging, RV measurements, and light curve modelling to show that \epic is a transiting planetary system, and not an EB or BEB. We discuss potential explanations for the striking fact that the two confirmed planets with the shortest orbital periods are both hosted by M-dwarfs.

\section*{Acknowledgements}

This paper includes data collected by the Kepler mission. Funding for the Kepler mission is provided by the NASA Science Mission directorate. Some of the data presented in this paper were obtained from the Mikulski Archive for Space Telescopes (MAST). STScI is operated by the Association of Universities for Research in Astronomy, Inc., under NASA contract NAS5-26555. Support for MAST for non-HST data is provided by the NASA Office of Space Science via grant NNX09AF08G and by other grants and contracts.

We greatly thank the NOT staff members for their precious support during the observations. Based on observations obtained with the Nordic Optical Telescope (NOT), operated on the island of La Palma jointly by Denmark, Finland, Iceland, Norway, and Sweden, in the Spanish Observatorio del Roque de los Muchachos (ORM) of the Instituto de Astrof\'isica de Canarias (IAC). The data presented here were obtained in part with ALFOSC, which is provided by the Instituto de Astrof\'isica de Andalucia (IAA) under a joint agreement with the University of Copenhagen and NOTSA. D.~Gandolfi gratefully acknowledges the financial support of the \emph{Programma Giovani Ricercatori -- Rita Levi Montalcini -- Rientro dei Cervelli (2012)} awarded by the Italian Ministry of Education, Universities and Research (MIUR).

Some of the data presented herein were obtained at the W.M. Keck Observatory, which is operated as a scientific partnership among the California Institute of Technology, the University of California and the National Aeronautics and Space Administration. The Observatory was made possible by the generous financial support of the W.M. Keck Foundation.

The WIYN/NESSI observations were conducted as part of an approved NOAO observing program (P.I. Livingston, proposal ID 2017A-0377). Data presented herein were obtained at the WIYN Observatory from telescope time allocated to NN-EXPLORE through the scientific
partnership of the National Aeronautics and Space Administration,
the National Science Foundation, and the National Optical Astronomy
Observatory. This work was supported by a NASA WIYN PI Data Award,
administered by the NASA Exoplanet Science Institute. NESSI was funded by the NASA Exoplanet Exploration Program and the NASA Ames Research Center. NESSI was built at the Ames Research Center by Steve B. Howell, Nic Scott, Elliott P. Horch, and Emmett Quigley.

This research has made use of NASA's Astrophysics Data System, the SIMBAD data base, operated at CDS, Strasbourg, France, the Exoplanet Orbit Database and the Exoplanet Data Explorer at exoplanets.org, and the Exoplanets Encyclopaedia at exoplanet.eu. We also used Astropy, a community-developed core Python package for Astronomy \citep{astropy}, and the DASCH project at Harvard, which is partially supported by NSF grants AST-0407380, AST-0909073, and AST-1313370.

The Pan-STARRS1 Surveys (PS1) have been made possible through contributions of the Institute for Astronomy, the University of Hawaii, the Pan-STARRS Project Office, the Max-Planck Society and its participating institutes, the Max Planck Institute for Astronomy, Heidelberg and the Max Planck Institute for Extraterrestrial Physics, Garching, The Johns Hopkins University, Durham University, the University of Edinburgh, Queen's University Belfast, the Harvard-Smithsonian Center for Astrophysics, the Las Cumbres Observatory Global Telescope Network Incorporated, the National Central University of Taiwan, the Space Telescope Science Institute, the National Aeronautics and Space Administration under Grant No. NNX08AR22G issued through the Planetary Science Division of the NASA Science Mission Directorate, the National Science Foundation under Grant No. AST-1238877, the University of Maryland, and Eotvos Lorand University (ELTE).

The Digitized Sky Surveys were produced at the Space Telescope Science Institute under U.S. Government grant NAG W-2166. The images of these surveys are based on photographic data obtained using the Oschin Schmidt Telescope on Palomar Mountain and the UK Schmidt Telescope. The plates were processed into the present compressed digital form with the permission of these institutions. The National Geographic Society - Palomar Observatory Sky Atlas (POSS-I) was made by the California Institute of Technology with grants from the National Geographic Society. The Second Palomar Observatory Sky Survey (POSS-II) was made by the California Institute of Technology with funds from the National Science Foundation, the National Geographic Society, the Sloan Foundation, the Samuel Oschin Foundation, and the Eastman Kodak Corporation.

Sz. Cs. acknowledges support from the Hungarian OTKA Grant K113117. Finally, we wish to thank the anonymous referee whose comments helped to improve the manuscript.

\bibliographystyle{mnras}
\bibliography{refs2} 

\begin{thebibliography}{}
\makeatletter
\relax
\def\mn@urlcharsother{\let\do\@makeother \do\$\do\&\do\#\do\^\do\_\do\%\do\~}
\def\mn@doi{\begingroup\mn@urlcharsother \@ifnextchar [ {\mn@doi@}
  {\mn@doi@[]}}
\def\mn@doi@[#1]#2{\def\@tempa{#1}\ifx\@tempa\@empty \href
  {http://dx.doi.org/#2} {doi:#2}\else \href {http://dx.doi.org/#2} {#1}\fi
  \endgroup}
\def\mn@eprint#1#2{\mn@eprint@#1:#2::\@nil}
\def\mn@eprint@arXiv#1{\href {http://arxiv.org/abs/#1} {{\tt arXiv:#1}}}
\def\mn@eprint@dblp#1{\href {http://dblp.uni-trier.de/rec/bibtex/#1.xml}
  {dblp:#1}}
\def\mn@eprint@#1:#2:#3:#4\@nil{\def\@tempa {#1}\def\@tempb {#2}\def\@tempc
  {#3}\ifx \@tempc \@empty \let \@tempc \@tempb \let \@tempb \@tempa \fi \ifx
  \@tempb \@empty \def\@tempb {arXiv}\fi \@ifundefined
  {mn@eprint@\@tempb}{\@tempb:\@tempc}{\expandafter \expandafter \csname
  mn@eprint@\@tempb\endcsname \expandafter{\@tempc}}}

\bibitem[\protect\citeauthoryear{{Adams}, {Jackson}  \& {Endl}}{{Adams}
  et~al.}{2016}]{Adams16}
{Adams} E.~R.,  {Jackson} B.,   {Endl} M.,  2016, \mn@doi [\aj]
  {10.3847/0004-6256/152/2/47}, \href
  {http://adsabs.harvard.edu/abs/2016AJ....152...47A} {152, 47}

\bibitem[\protect\citeauthoryear{{Astropy Collaboration} et~al.,}{{Astropy
  Collaboration} et~al.}{2013}]{astropy}
{Astropy Collaboration} et~al., 2013, \mn@doi [\aap]
  {10.1051/0004-6361/201322068}, \href
  {http://adsabs.harvard.edu/abs/2013A%26A...558A..33A} {558, A33}

\bibitem[\protect\citeauthoryear{{Barker} \& {Ogilvie}}{{Barker} \&
  {Ogilvie}}{2009}]{Barker+Ogilvie}
{Barker} A.~J.,  {Ogilvie} G.~I.,  2009, \mn@doi [\mnras]
  {10.1111/j.1365-2966.2009.14694.x}, \href
  {http://adsabs.harvard.edu/abs/2009MNRAS.395.2268B} {395, 2268}

\bibitem[\protect\citeauthoryear{{Bochanski}, {West}, {Hawley}  \&
  {Covey}}{{Bochanski} et~al.}{2007}]{Bochanski2007}
{Bochanski} J.~J.,  {West} A.~A.,  {Hawley} S.~L.,   {Covey} K.~R.,  2007,
  \mn@doi [\aj] {10.1086/510240}, \href
  {http://adsabs.harvard.edu/abs/2007AJ....133..531B} {133, 531}

\bibitem[\protect\citeauthoryear{{Cabrera}, {Csizmadia}, {Erikson}, {Rauer}  \&
  {Kirste}}{{Cabrera} et~al.}{2012}]{DST}
{Cabrera} J.,  {Csizmadia} S.,  {Erikson} A.,  {Rauer} H.,   {Kirste} S.,
  2012, \mn@doi [\aap] {10.1051/0004-6361/201219337}, \href
  {http://adsabs.harvard.edu/abs/2012A%26A...548A..44C} {548, A44}

\bibitem[\protect\citeauthoryear{{Castelli} \& {Kurucz}}{{Castelli} \&
  {Kurucz}}{2004}]{Castelli2004}
{Castelli} F.,  {Kurucz} R.~L.,  2004, preprint, \href
  {http://adsabs.harvard.edu/abs/2004astro.ph..5087C} {} (\mn@eprint {arXiv}
  {0405087})

\bibitem[\protect\citeauthoryear{{Csizmadia} et~al.,}{{Csizmadia}
  et~al.}{2015}]{Szilard_BD}
{Csizmadia} S.,  et~al., 2015, \mn@doi [\aap] {10.1051/0004-6361/201526763},
  \href {http://adsabs.harvard.edu/abs/2015A%26A...584A..13C} {584, A13}

\bibitem[\protect\citeauthoryear{{Frandsen} \& {Lindberg}}{{Frandsen} \&
  {Lindberg}}{1999}]{Frandsen1999}
{Frandsen} S.,  {Lindberg} B.,  1999, in {Karttunen} H.,  {Piirola} V.,  eds,
  Astrophysics with the NOT. p.~71

\bibitem[\protect\citeauthoryear{{Gandolfi} et~al.,}{{Gandolfi}
  et~al.}{2008}]{Gandolfi2008}
{Gandolfi} D.,  et~al., 2008, \mn@doi [\apj] {10.1086/591729}, \href
  {http://adsabs.harvard.edu/abs/2008ApJ...687.1303G} {687, 1303}

\bibitem[\protect\citeauthoryear{{Gandolfi} et~al.,}{{Gandolfi}
  et~al.}{2015}]{Gandolfi2015}
{Gandolfi} D.,  et~al., 2015, \mn@doi [\aap] {10.1051/0004-6361/201425062},
  \href {http://adsabs.harvard.edu/abs/2015A%26A...576A..11G} {576, A11}

\bibitem[\protect\citeauthoryear{{Girardi}, {Groenewegen}, {Hatziminaoglou}  \&
  {da Costa}}{{Girardi} et~al.}{2005}]{trilegal}
{Girardi} L.,  {Groenewegen} M.~A.~T.,  {Hatziminaoglou} E.,   {da Costa} L.,
  2005, \mn@doi [\aap] {10.1051/0004-6361:20042352}, \href
  {http://cdsads.u-strasbg.fr/abs/2005A%26A...436..895G} {436, 895}

\bibitem[\protect\citeauthoryear{{Hawley} et~al.,}{{Hawley}
  et~al.}{2002}]{Hawley2002}
{Hawley} S.~L.,  et~al., 2002, \mn@doi [\aj] {10.1086/340697}, \href
  {http://adsabs.harvard.edu/abs/2002AJ....123.3409H} {123, 3409}

\bibitem[\protect\citeauthoryear{{Hellier} et~al.,}{{Hellier}
  et~al.}{2011}]{w43}
{Hellier} C.,  et~al., 2011, \mn@doi [\aap] {10.1051/0004-6361/201117081},
  \href {http://esoads.eso.org/abs/2011A%26A...535L...7H} {535, L7}

\bibitem[\protect\citeauthoryear{{Hirano} et~al.,}{{Hirano}
  et~al.}{2016}]{Hirano16}
{Hirano} T.,  et~al., 2016, \mn@doi [\apj] {10.3847/0004-637X/820/1/41}, \href
  {http://adsabs.harvard.edu/abs/2016ApJ...820...41H} {820, 41}

\bibitem[\protect\citeauthoryear{{Howard} et~al.,}{{Howard}
  et~al.}{2013}]{kepler78_2}
{Howard} A.~W.,  et~al., 2013, \mn@doi [\nat] {10.1038/nature12767}, \href
  {http://adsabs.harvard.edu/abs/2013Natur.503..381H} {503, 381}

\bibitem[\protect\citeauthoryear{{Howell}, {Everett}, {Sherry}, {Horch}  \&
  {Ciardi}}{{Howell} et~al.}{2011}]{speckle}
{Howell} S.~B.,  {Everett} M.~E.,  {Sherry} W.,  {Horch} E.,   {Ciardi} D.~R.,
  2011, \mn@doi [\aj] {10.1088/0004-6256/142/1/19}, \href
  {http://adsabs.harvard.edu/abs/2011AJ....142...19H} {142, 19}

\bibitem[\protect\citeauthoryear{{Jackson}, {Greenberg}  \& {Barnes}}{{Jackson}
  et~al.}{2008}]{Jackson08}
{Jackson} B.,  {Greenberg} R.,   {Barnes} R.,  2008, \mn@doi [ApJ]
  {10.1086/529187}, \href {http://adsabs.harvard.edu/abs/2008ApJ...678.1396J}
  {678, 1396}

\bibitem[\protect\citeauthoryear{{Kipping}}{{Kipping}}{2010}]{Kipping10}
{Kipping} D.~M.,  2010, \mn@doi [\mnras] {10.1111/j.1365-2966.2010.17242.x},
  \href {http://adsabs.harvard.edu/abs/2010MNRAS.408.1758K} {408, 1758}

\bibitem[\protect\citeauthoryear{{Kirk} et~al.,}{{Kirk}
  et~al.}{2016}]{Kep_binary_cat}
{Kirk} B.,  et~al., 2016, \mn@doi [\aj] {10.3847/0004-6256/151/3/68}, \href
  {http://adsabs.harvard.edu/abs/2016AJ....151...68K} {151, 68}

\bibitem[\protect\citeauthoryear{{Kobayashi} et~al.,}{{Kobayashi}
  et~al.}{2000}]{Subaru_IRCS}
{Kobayashi} N.,  et~al., 2000, in {Iye} M.,  {Moorwood} A.~F.,  eds,  \procspie
  Vol. 4008, Optical and IR Telescope Instrumentation and Detectors. pp
  1056--1066, \mn@doi{10.1117/12.395423}

\bibitem[\protect\citeauthoryear{{Kotani} et~al.,}{{Kotani} et~al.}{2014}]{IRD}
{Kotani} T.,  et~al., 2014, in Ground-based and Airborne Instrumentation for
  Astronomy V. p. 914714, \mn@doi{10.1117/12.2055075}

\bibitem[\protect\citeauthoryear{{Le Borgne} et~al.,}{{Le Borgne}
  et~al.}{2003}]{LeBorgne2003}
{Le Borgne} J.-F.,  et~al., 2003, \mn@doi [\aap] {10.1051/0004-6361:20030243},
  \href {http://adsabs.harvard.edu/abs/2003A%26A...402..433L} {402, 433}

\bibitem[\protect\citeauthoryear{{L{\'e}ger} et~al.,}{{L{\'e}ger}
  et~al.}{2009}]{corot7_2}
{L{\'e}ger} A.,  et~al., 2009, \mn@doi [\aap] {10.1051/0004-6361/200911933},
  \href {http://adsabs.harvard.edu/abs/2009A%26A...506..287L} {506, 287}

\bibitem[\protect\citeauthoryear{{Mandel} \& {Agol}}{{Mandel} \&
  {Agol}}{2002}]{M&A}
{Mandel} K.,  {Agol} E.,  2002, \mn@doi [ApJ] {10.1086/345520}, \href
  {http://adsabs.harvard.edu/abs/2002ApJ...580L.171M} {580, L171}

\bibitem[\protect\citeauthoryear{{Mann}, {Feiden}, {Gaidos}, {Boyajian}  \&
  {von Braun}}{{Mann} et~al.}{2015}]{Mann15}
{Mann} A.~W.,  {Feiden} G.~A.,  {Gaidos} E.,  {Boyajian} T.,   {von Braun} K.,
  2015, \mn@doi [\apj] {10.1088/0004-637X/804/1/64}, \href
  {http://adsabs.harvard.edu/abs/2015ApJ...804...64M} {804, 64}

\bibitem[\protect\citeauthoryear{{Mart{\'{\i}}n}, {Delfosse}, {Basri},
  {Goldman}, {Forveille}  \& {Zapatero Osorio}}{{Mart{\'{\i}}n}
  et~al.}{1999}]{Martin1999}
{Mart{\'{\i}}n} E.~L.,  {Delfosse} X.,  {Basri} G.,  {Goldman} B.,  {Forveille}
  T.,   {Zapatero Osorio} M.~R.,  1999, \mn@doi [\aj] {10.1086/301107}, \href
  {http://adsabs.harvard.edu/abs/1999AJ....118.2466M} {118, 2466}

\bibitem[\protect\citeauthoryear{{Matijevi{\v c}}, {Pr{\v s}a}, {Orosz},
  {Welsh}, {Bloemen}  \& {Barclay}}{{Matijevi{\v c}}
  et~al.}{2012}]{Matijevic12}
{Matijevi{\v c}} G.,  {Pr{\v s}a} A.,  {Orosz} J.~A.,  {Welsh} W.~F.,
  {Bloemen} S.,   {Barclay} T.,  2012, \mn@doi [\aj]
  {10.1088/0004-6256/143/5/123}, \href
  {http://adsabs.harvard.edu/abs/2012AJ....143..123M} {143, 123}

\bibitem[\protect\citeauthoryear{{Morris} \& {Naftilan}}{{Morris} \&
  {Naftilan}}{1993}]{Morris93}
{Morris} S.~L.,  {Naftilan} S.~A.,  1993, \mn@doi [\apj] {10.1086/173488},
  \href {http://adsabs.harvard.edu/abs/1993ApJ...419..344M} {419, 344}

\bibitem[\protect\citeauthoryear{{Muirhead} et~al.,}{{Muirhead}
  et~al.}{2013}]{Muirhead13}
{Muirhead} P.~S.,  et~al., 2013, \mn@doi [\apj] {10.1088/0004-637X/767/2/111},
  \href {http://cdsads.u-strasbg.fr/abs/2013ApJ...767..111M} {767, 111}

\bibitem[\protect\citeauthoryear{{Ochsenbein}, {Bauer}  \&
  {Marcout}}{{Ochsenbein} et~al.}{2000}]{Vizier}
{Ochsenbein} F.,  {Bauer} P.,   {Marcout} J.,  2000, \mn@doi [\aaps]
  {10.1051/aas:2000169}, \href
  {http://adsabs.harvard.edu/abs/2000A%26AS..143...23O} {143, 23}

\bibitem[\protect\citeauthoryear{{Ofir} \& {Dreizler}}{{Ofir} \&
  {Dreizler}}{2013}]{Ofir_Dreizler_13}
{Ofir} A.,  {Dreizler} S.,  2013, \mn@doi [\aap] {10.1051/0004-6361/201219877},
  \href {http://adsabs.harvard.edu/abs/2013A%26A...555A..58O} {555, A58}

\bibitem[\protect\citeauthoryear{{Oke}}{{Oke}}{1990}]{Oke90}
{Oke} J.~B.,  1990, \mn@doi [\aj] {10.1086/115444}, \href
  {http://adsabs.harvard.edu/abs/1990AJ.....99.1621O} {99, 1621}

\bibitem[\protect\citeauthoryear{{Pecaut} \& {Mamajek}}{{Pecaut} \&
  {Mamajek}}{2013}]{Pecaut_Mamajek}
{Pecaut} M.~J.,  {Mamajek} E.~E.,  2013, \mn@doi [\apjs]
  {10.1088/0067-0049/208/1/9}, \href
  {http://adsabs.harvard.edu/abs/2013ApJS..208....9P} {208, 9}

\bibitem[\protect\citeauthoryear{{Pepe} et~al.,}{{Pepe}
  et~al.}{2013}]{kepler78_3}
{Pepe} F.,  et~al., 2013, \mn@doi [\nat] {10.1038/nature12768}, \href
  {http://adsabs.harvard.edu/abs/2013Natur.503..377P} {503, 377}

\bibitem[\protect\citeauthoryear{{Queloz} et~al.,}{{Queloz}
  et~al.}{2009}]{corot7_1}
{Queloz} D.,  et~al., 2009, \mn@doi [\aap] {10.1051/0004-6361/200913096}, \href
  {http://adsabs.harvard.edu/abs/2009A%26A...506..303Q} {506, 303}

\bibitem[\protect\citeauthoryear{{Quirrenbach} et~al.,}{{Quirrenbach}
  et~al.}{2010}]{carmenes}
{Quirrenbach} A.,  et~al., 2010, in Ground-based and Airborne Instrumentation
  for Astronomy III. p. 773513, \mn@doi{10.1117/12.857777}

\bibitem[\protect\citeauthoryear{{Rappaport}, {Sanchis-Ojeda}, {Rogers},
  {Levine}  \& {Winn}}{{Rappaport} et~al.}{2013}]{KOI1843}
{Rappaport} S.,  {Sanchis-Ojeda} R.,  {Rogers} L.~A.,  {Levine} A.,   {Winn}
  J.~N.,  2013, \mn@doi [\apjl] {10.1088/2041-8205/773/1/L15}, \href
  {http://adsabs.harvard.edu/abs/2013ApJ...773L..15R} {773, L15}

\bibitem[\protect\citeauthoryear{{Reynolds} \& {Summers}}{{Reynolds} \&
  {Summers}}{1969}]{Reynolds69}
{Reynolds} R.~T.,  {Summers} A.~L.,  1969, \mn@doi [\jgr]
  {10.1029/JB074i010p02494}, \href
  {http://adsabs.harvard.edu/abs/1969JGR....74.2494R} {74, 2494}

\bibitem[\protect\citeauthoryear{{Sanchis-Ojeda}, {Rappaport}, {Winn},
  {Levine}, {Kotson}, {Latham}  \& {Buchhave}}{{Sanchis-Ojeda}
  et~al.}{2013}]{kepler78}
{Sanchis-Ojeda} R.,  {Rappaport} S.,  {Winn} J.~N.,  {Levine} A.,  {Kotson}
  M.~C.,  {Latham} D.~W.,   {Buchhave} L.~A.,  2013, \mn@doi [\apj]
  {10.1088/0004-637X/774/1/54}, \href
  {http://adsabs.harvard.edu/abs/2013ApJ...774...54S} {774, 54}

\bibitem[\protect\citeauthoryear{{Sanchis-Ojeda}, {Rappaport}, {Winn},
  {Kotson}, {Levine}  \& {El Mellah}}{{Sanchis-Ojeda}
  et~al.}{2014}]{Sanchis-Ojeda14}
{Sanchis-Ojeda} R.,  {Rappaport} S.,  {Winn} J.~N.,  {Kotson} M.~C.,  {Levine}
  A.,   {El Mellah} I.,  2014, \mn@doi [\apj] {10.1088/0004-637X/787/1/47},
  \href {http://adsabs.harvard.edu/abs/2014ApJ...787...47S} {787, 47}

\bibitem[\protect\citeauthoryear{{Seager}, {Kuchner}, {Hier-Majumder}  \&
  {Militzer}}{{Seager} et~al.}{2007}]{Seager07}
{Seager} S.,  {Kuchner} M.,  {Hier-Majumder} C.~A.,   {Militzer} B.,  2007,
  \mn@doi [\apj] {10.1086/521346}, \href
  {http://adsabs.harvard.edu/abs/2007ApJ...669.1279S} {669, 1279}

\bibitem[\protect\citeauthoryear{{Shan}, {Johnson}  \& {Morton}}{{Shan}
  et~al.}{2015}]{Shan15}
{Shan} Y.,  {Johnson} J.~A.,   {Morton} T.~D.,  2015, \mn@doi [\apj]
  {10.1088/0004-637X/813/1/75}, \href
  {http://adsabs.harvard.edu/abs/2015ApJ...813...75S} {813, 75}

\bibitem[\protect\citeauthoryear{{Sing}}{{Sing}}{2010}]{Sing09}
{Sing} D.~K.,  2010, \mn@doi [\aap] {10.1051/0004-6361/200913675}, \href
  {http://adsabs.harvard.edu/abs/2010A%26A...510A..21S} {510, A21}

\bibitem[\protect\citeauthoryear{{Smith} et~al.,}{{Smith} et~al.}{2017}]{K299}
{Smith} A.~M.~S.,  et~al., 2017, \mn@doi [\mnras] {10.1093/mnras/stw2487},
  \href {http://adsabs.harvard.edu/abs/2017MNRAS.464.2708S} {464, 2708}

\bibitem[\protect\citeauthoryear{{Telting} et~al.,}{{Telting}
  et~al.}{2014}]{Telting2014}
{Telting} J.~H.,  et~al., 2014, \mn@doi [Astronomische Nachrichten]
  {10.1002/asna.201312007}, \href
  {http://adsabs.harvard.edu/abs/2014AN....335...41T} {335, 41}

\bibitem[\protect\citeauthoryear{{Tokovinin}}{{Tokovinin}}{2008}]{Tokovinin08}
{Tokovinin} A.,  2008, \mn@doi [\mnras] {10.1111/j.1365-2966.2008.13613.x},
  \href {http://adsabs.harvard.edu/abs/2008MNRAS.389..925T} {389, 925}

\bibitem[\protect\citeauthoryear{{Vanderburg} \& {Johnson}}{{Vanderburg} \&
  {Johnson}}{2014}]{vburg}
{Vanderburg} A.,  {Johnson} J.~A.,  2014, \mn@doi [\pasp] {10.1086/678764},
  \href {http://adsabs.harvard.edu/abs/2014PASP..126..948V} {126, 948}

\bibitem[\protect\citeauthoryear{{Winn} et~al.,}{{Winn}
  et~al.}{2017}]{Winn_USP}
{Winn} J.~N.,  et~al., 2017, AJ, in press (arXiv:1704.00203), \href
  {http://adsabs.harvard.edu/abs/2017arXiv170400203W} {}

\bibitem[\protect\citeauthoryear{{Yee}, {Petigura}  \& {von Braun}}{{Yee}
  et~al.}{2017}]{Specmatch-emp}
{Yee} S.~W.,  {Petigura} E.~A.,   {von Braun} K.,  2017, \mn@doi [\apj]
  {10.3847/1538-4357/836/1/77}, \href
  {http://adsabs.harvard.edu/abs/2017ApJ...836...77Y} {836, 77}

\bibitem[\protect\citeauthoryear{{Zacharias}, {Finch}  \&
  {Frouard}}{{Zacharias} et~al.}{2017}]{UCAC5}
{Zacharias} N.,  {Finch} C.,   {Frouard} J.,  2017, \mn@doi [\aj]
  {10.3847/1538-3881/aa6196}, \href
  {http://adsabs.harvard.edu/abs/2017AJ....153..166Z} {153, 166}

\makeatother
\end{thebibliography}






\bsp	
\label{lastpage}
\end{document}